\documentclass[apj]{emulateapj}
\pdfoutput = 1
\usepackage{graphicx, subfigure}
\usepackage{amsmath, natbib, listings}
\usepackage[bottom, hang, flushmargin]{footmisc}
\usepackage[urlcolor=magenta,citecolor=green,linkcolor=red]{hyperref}
\usepackage{appendix}
\usepackage{rotate}
\usepackage{adjustbox}

\usepackage{amsmath, cancel, amssymb}
\usepackage{bm}

\def\jcap{\ref@jnl{J. Cosmology Astropart. Phys.}}%

\def \gsim { \lower .75ex \hbox{$\sim$} \llap{\raise .27ex \hbox{$>$}} }
\def \lsim { \lower .75ex \hbox{$\sim$} \llap{\raise .27ex \hbox{$<$}} }
\def \deg {^{\circ}}

\def \Spitzer  {{\it Spitzer \,}}

\newcommand{\sqdeg}{deg$^{2}$ }

\begin{document}

\shorttitle{SpIES: Survey Overview}
\shortauthors{}

\title{S\MakeLowercase{p}IES: The \emph{Spitzer} IRAC Equatorial Survey}

\author{
  John D. Timlin\altaffilmark{1,$\star$},
  Nicholas P. Ross\altaffilmark{1,2}, 
 Gordon T. Richards\altaffilmark{1},
 Mark Lacy\altaffilmark{3},
 Erin L. Ryan\altaffilmark{4},
 Robert B. Stone\altaffilmark{1},
Franz E. Bauer\altaffilmark{5,6,7},
W. N. Brandt\altaffilmark{8,9,10},
Xiaohui Fan\altaffilmark{11},
Eilat Glikman\altaffilmark{12},
Daryl Haggard\altaffilmark{13},
Linhua Jiang\altaffilmark{14},
Stephanie M. LaMassa\altaffilmark{15},
Yen-Ting Lin\altaffilmark{16},
Martin Makler\altaffilmark{17},
Peregrine McGehee\altaffilmark{18},
Adam D. Myers\altaffilmark{19},
Donald P. Schneider\altaffilmark{8,9},
C. Megan Urry\altaffilmark{20},
Edward J. Wollack\altaffilmark{21},
Nadia L. Zakamska\altaffilmark{22}
}

\altaffiltext{$\star$}{For correspondence regarding this article, please write to J.~D. Timlin: \email{john.d.timlin@drexel.edu}}
\altaffiltext{1}{Department of Physics, Drexel University, 3141 Chestnut Street, Philadelphia, PA 19104, U.S.A}
\altaffiltext{2}{Institute for Astronomy, University of Edinburgh, Royal Observatory, Edinburgh, EH9 3HJ, U.K.}
\altaffiltext{3}{National Radio Astronomy Observatory, 520 Edgemont Road, Charlottesville, VA 22903, U.S.A}
\altaffiltext{4}{University of Maryland Department of Astronomy, College Park, MD 20742, U.S.A}
\altaffiltext{5}{Instituto de Astrof\'{\i}sica, Facultad de F\'{i}sica, Pontificia Universidad Cat\'{o}lica de Chile, Casilla 306, Santiago 22, Chile} 
\altaffiltext{6}{Millennium Institute of Astrophysics, MAS, Nuncio Monse\~{n}or S\'{o}tero Sanz 100, Providencia, Santiago de Chile} 
\altaffiltext{7}{Space Science Institute, 4750 Walnut Street, Suite 205, Boulder, Colorado 80301} 
\altaffiltext{8}{Department of Astronomy \& Astrophysics, 525 Davey Lab, The Pennsylvania State University, University Park, PA 16802, USA}
\altaffiltext{9}{Institute for Gravitation and the Cosmos, The Pennsylvania State University, University Park, PA 16802, USA}
\altaffiltext{10}{Department of Physics, 104 Davey Lab, The Pennsylvania State University, University Park, PA 16802, USA}
\altaffiltext{11}{Steward Observatory, University of Arizona, Tucson, AZ 85721, USA}
\altaffiltext{12}{Department of Physics, Middlebury College, Middlebury, VT 05753, USA}
\altaffiltext{13}{Department of Physics and Astronomy, Amherst College, Amherst, MA 01002-5000, USA}
\altaffiltext{14}{Kavli Institute for Astronomy and Astrophysics, Peking University, Beijing 100871, China}
\altaffiltext{15}{NPP Fellow, NASA GSFC, Greenbelt, MD, 20771}
\altaffiltext{16}{Institute of Astronomy and Astrophysics, Academia Sinica, Taipei 106, Taiwan}
\altaffiltext{17}{Centro Brasileiro de Pesquisas F\'isicas, Rua Dr. Xavier Sigaud 150, CEP 22290-180, Rio de Janeiro, RJ, Brazil}
\altaffiltext{18}{IPAC, 1200 E. California Blvd, Pasadena, CA 91125}
\altaffiltext{19}{Department of Physics and Astronomy, University of Wyoming, 1000 University Ave., Laramie, WY, 82071, USA}
\altaffiltext{20}{Yale Center for Astronomy and Astrophysics, Yale University, Physics Department, PO Box 208120, New Haven, CT, 06520-8120, USA}
\altaffiltext{21}{NASA Goddard Space Flight Center, Greenbelt, MD 20771}
\altaffiltext{22}{Department of Physics and Astronomy, Johns Hopkins University, Bloomberg Center, 3400 N. Charles St., Baltimore, MD 21218, USA}

\date{\today}

\begin{abstract}
We describe the first data release from the \emph{Spitzer}-IRAC Equatorial Survey (SpIES); a large-area survey of $\sim$115 deg$^{2}$ in the Equatorial SDSS Stripe 82 field using \emph{Spitzer} during its `warm' mission phase. SpIES was designed to probe sufficient volume to perform measurements of quasar clustering and the luminosity function at $z \geq$ 3 to test various models for ``feedback" from active galactic nuclei (AGN). Additionally, the wide range of available multi-wavelength, multi-epoch ancillary data enables SpIES to identify both high-redshift ($z\geq$ 5) quasars as well as obscured quasars missed by optical surveys. SpIES achieves 5$\sigma$ depths of 6.13 $\mu$Jy (21.93 AB magnitude) and 5.75 $\mu$Jy (22.0 AB magnitude) at 3.6 and 4.5 microns, respectively---depths significantly fainter than \emph{WISE}. We show that the SpIES survey recovers a much larger fraction of spectroscopically-confirmed quasars ($\sim$98$\%$) in Stripe 82 than are recovered by \emph{WISE} ($\sim$55$\%$). This depth is especially powerful at high-redshift ($z\ge 3.5$), where SpIES recovers 94$\%$ of confirmed quasars, whereas \emph{WISE} only recovers 25$\%$. Here we define the SpIES survey parameters and describe the image processing, source extraction, and catalog production methods used to analyze the SpIES data. In addition to this survey paper, we release 234 images created by the SpIES team and three detection catalogs: a 3.6\,$\mu$m-only detection catalog containing $\sim$6.1 million sources, a 4.5\,$\mu$m-only detection catalog containing $\sim$6.5 million sources, and a dual-band detection catalog containing $\sim$5.4 million sources.
\end{abstract}


\maketitle

\section{Introduction}
\label{sec:intro}
The \Spitzer Space Telescope \citep{Werner2004} has been paramount in understanding the Universe at mid-infrared wavelengths. During its primary mission, \Spitzer observed at 3.6, 4.5, 5.8, and 8.0 $\mu$m using the Infrared Array Camera (IRAC; \citealt{Fazio2004b}), at 24, 70, and 160 $\mu$m using the Multiband Imaging Photometer for \Spitzer (MIPS; \citealt{Rieke2004}) camera, and had a dedicated infrared spectrograph (IRS; \citealt{Houck2004}) covering wavelengths from 5.3 to 38 $\mu$m. Since the exhaustion of its cryogen in 2009, \emph{Spitzer} has run its `warm' mission phase, taking images with the two shortest IRAC passbands (3.6 and 4.5 $\mu$m). 

\Spitzer IRAC has been a valuable tool in the creation of deep, relatively small area surveys through campaigns like the $\sim$2 \sqdeg Spitzer-COSMOS survey (S-COSMOS; \citealt{Sanders2007}) and the $\sim$10 \sqdeg \Spitzer Deep, Wide-field Survey (SDWFS; \citealt{Ashby2009}) utilizing all four of the IRAC bands. \emph{Spitzer} continues to delve deeper in its `warm' phase with the IRAC ultradeep filed (IUDF; \citealt{Labbe2015}), the $\sim$1.2 \sqdeg \emph{Spitzer} Large Area Survey with Hyper-Suprime-Cam (SPLASH; \citealt{Steinhardt2014}), and the $\sim$18 \sqdeg \emph{Spitzer} Extragalactic Representative Volume Survey (SERVS; \citealt{Mauduit2012}).

Despite having a relatively small 5$\farcm$2$\times$5$\farcm$2 field of view (FOV), IRAC has also effectively and efficiently run larger-area programs throughout its lifetime such as the $\sim$65 deg$^2$ SIRTF Wide-Area Infrared Extragalactic Survey (SWIRE; \citealt{Lonsdale2003}). Recently, \emph{Spitzer} has made an effort to run larger-area surveys in the `warm' phase with the $\sim$26 deg$^2$ \emph{Spitzer}-HETDEX Exploratory Large Area \citep{Papovich2016} and the $\sim$94 deg$^2$ \emph{Spitzer} South Pole Telescope Deep Field (SSDF; \citealt{Ashby2013}) mission which, until now, had the largest area of any \Spitzer survey. 

These large-area campaigns are made possible by the IRAC mapping mode strategy, which aligns the arrays on a positional grid, allowing observations to overlap through successive motions in the grid. This approach differs from other observing strategies, many of which forced the telescope to slew to a single position multiple times to observe the same location on the sky in a different channel (see Section 3.2 of the IRAC Instrument Handbook\footnote{\href{IRAC}{http://irsa.ipac.caltech.edu/data/SPITZER\\/docs/irac/iracinstrumenthandbook/}\label{IRAC}}). Mapping mode decreases slew time, allowing for larger area surveys to be performed while still reaching interesting flux limits. 

{\emph{Spitzer}} is not the only telescope performing large area, mid-infrared observations of the Universe. The Wide-field Infrared Survey Explorer ({\emph{WISE}}; \citealt{Wright2010}) telescope has been mapping the entire sky in four channels, two of which have nearly the same wavelength as `warm' \emph{Spitzer} (3.4 and 4.6 $\mu$m). While {\emph{WISE}} covers essentially the entire sky, it lacks both the depth and the spatial resolution that \Spitzer IRAC surveys can achieve.  

\begin{deluxetable}{ll}[ht!]
\tablecolumns{2}
\tablewidth{0pt}
\tablecaption{The \Spitzer IRAC Equatorial Survey (S\MakeLowercase{p}IES) key parameters}
\tablehead{
\colhead{Parameter} & \colhead{Value}}
\startdata
Imaging                                       & IRAC Ch1 and Ch2   \\ 
Wavelength                                  & 3.6 and 4.5 $\mu$m   \\
Area$^{a}$                                             & $\sim$115 deg$^{2}$ 	                    \\
No. of IRAC pointings                         & $\sim$70,000             \\
Exposure Time at each pointing  & 60s          \\ 
Total Observation Time               & 820hr       \\   
Typical Zodiacal Background       & $0.09-0.23$ MJy sr$^{-1}$ \\
IRAC PSF FWHM$^{b}$			& 1$\farcs$95, 2$\farcs$02		\\
Total number of objects$^{c}$     & $\sim$5,400,000 \\
Limiting AB Magnitude$^{d}$ (5$\sigma$)	& 21.93, 22.0 \\
Data URL:   &      \\
\multicolumn{2}{c}{\url{http://www.physics.drexel.edu/~gtr/spies/}}
\enddata
\tablecomments{$^{a}$ Total survey area covered by both detectors. The area covered by a single detector decreases due to their separation on IRAC (details in Section \ref{sec:Obs_Strat}).  $^{b}$5$\sigma$ dual-band detection catalog (see Section \ref{catalog}). $^{c}$Total number of objects in the dual-band catalog. $^{d}$Values are for the 3.6\,$\mu$m, 4.5\,$\mu$m detectors.}
\label{tab:keyparams}
\end{deluxetable}

In this paper, we describe the \Spitzer IRAC Equatorial Survey (SpIES) parameters and catalogs. SpIES mapped a large portion of the Sloan Digital Sky Survey (SDSS; \citealt{York2000}) equatorial S82 field (\citealt{Stoughton2002}; \citealt{Annis2014}; \citealt{Jiang2014}), utilizing the \Spitzer 3.6 and 4.5 $\mu$m bands (often referred to as Ch1 and Ch2 respectively). Collecting $\sim$115 \sqdeg over $\sim$820 hours, SpIES is the largest area \Spitzer survey, probing to depths comparable to SWIRE. Table \ref{tab:keyparams} contains the key parameters of SpIES such as the wavelengths and point spread function of IRAC, along with the observation times, area, and depth of the SpIES survey. With this release, we present three SpIES source catalogs consisting of $\sim$6.1 million objects detected only at 3.6\,$\mu$m, $\sim$6.6 million objects detected only at 4.5\,$\mu$m, and a dual-band detection catalog which contains $\sim$5.4 million detections in both bands. We also release the images generated by the SpIES team used to build the catalogs described herein.

The combined depth and area of the SpIES, along with the wealth of multi-wavelength, multi-epoch ancillary imaging and spectroscopic data on Stripe 82 (S82; \citealt{Stoughton2002}; \citealt{Annis2014}; \citealt{Jiang2014}), make it a powerful tool for addressing a wide range of topics in contemporary astrophysics. In particular, we seek to use the data to: probe the population of obscured quasars at high redshift (e.g.,\ \citealt{Alexandroff2013, Glikman2013, Assef2015}); use high-redshift unobscured quasars to investigate how quasar feedback contributes to galaxy evolution (e.g.,\ \citealt{Hopkins2007a}; \citealt{White2012}); improve the removal of foreground objects from maps of the cosmic microwave background \citep{Wang2006}; better constrain the stellar masses of Lyman Break Galaxies (e.g.,\ \citealt{Daddi2007}); improve stellar population modeling for hosts of supernovae (e.g.,\ \citealt{Sullivan2010, Fox2015}); and enable discovery of cool stars (e.g.,\ \citealt{Lucas2010}).

\begin{figure*}[!h]\label{S82footprint}
  \begin{adjustbox}{addcode={\begin{minipage}{\width}}{\caption{%
Top: We show the SpIES coverage area (yellow and purple rectangles) atop the 100$\mu$m IRAS dust map \citep{Schlegel1998} of the full SDSS Stripe 82 region (white box). Many different surveys have covered this region of the sky and overlap with SpIES.  Displayed are the HeLMS (green box) and HeRS (light blue) survey footprints \citep{Oliver2012, Viero2014}, the regions observed by \emph{XMM-Newton} (yellow and orange circles) and \emph{Chandra} (red circles were observed with the ACIS-S arrays and blue circles with the ACIS-I arrays; \citealt{LaMassa2013b, LaMassa2013c}),  the VLA (green scallop) from \citet{Hodge2011}, and the SHELA  observations (orange boxes) by \citet{Papovich2016}, as a few examples of many surveys that cover the S82 region. More details about other surveys on S82 can be found in Table \ref{Stripe82_data}. Bottom: Detailed SpIES 3.6\,$\mu$m (yellow) and 4.5\,$\mu$m (purple) coverage of Stripe 82 along with SHELA coverage (orange). Both panels are centered on $\delta$=0  and $\alpha$ values are given in J2000 degrees.
      }\end{minipage}},rotate=90,center}
      \includegraphics[scale=.5]{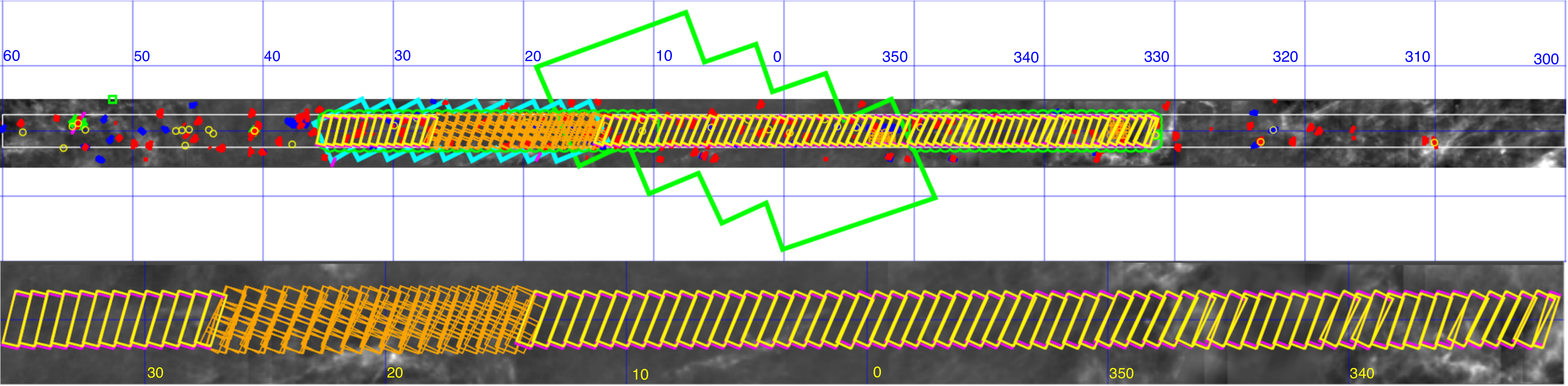}
  \end{adjustbox}
\end{figure*}

We begin our discussion by describing the existing data covering the S82 footprint in Section \ref{Stripe82}, followed by the \Spitzer observation strategy used for SpIES in Section \ref{sec:Obs_Strat}. We discuss the data products from \Spitzer and our image stacking process in Section \ref{sec:DatRed}. The SpIES catalogs are described Section \ref{catalog}, which includes source extraction techniques, photometric errors, and astrometric reliability. This section also discusses the completeness, number counts, and depth of the SpIES detection catalog. Finally, in Section \ref{sec:sourcematch}, we match SpIES objects to various quasar catalogs to test the SpIES recovery fraction of high-redshift quasars. We also provide a summary of the SpIES survey and links to the data products in Appendix \ref{AppendA} 

We calculate magnitudes on the AB scale, which has a flux density zeropoint of 3631Jy \citep{Oke1983}. These are denoted as [3.6] and [4.5], respectively. Conversion to Vega magnitudes is given by [3.6]$-$2.779 and [4.5]$-$3.264, respectively (calculated using the Vega zeropoint flux density values of 280.9 Jy at 3.6\,$\mu$m and 179.7 Jy at 4.5\,$\mu$m from Table 4.1 in the IRAC Handbook\footref{IRAC}).


\section{The Stripe 82 Region} \label{Stripe82}
The observational goal of the SpIES project was to map S82 in order to provide a suitably large ``laboratory" in which to conduct the types of experiments that involve rare objects, as noted above.  S82 is located on the Celestial Equator spanning a range of $-60\deg \leq \alpha \leq 60\deg$ and $-1.25\deg \leq \delta \leq 1.25\deg$. The SpIES observations cover approximately one third of this region centered on $\delta = 0\deg$ and spanning the range from $-30\deg \leq \alpha \leq 35\deg$, with a break in coverage between $13.9\deg\le \alpha \le 27.2\deg$ where deeper IRAC data exists from the SHELA \citep{Papovich2016} survey. Within those RA limits, SpIES completely covers S82 from $-0.85\deg \leq \delta \leq 0.85\deg$ with irregular coverage outside of that declination range due to the orientation of observations (see Figure \ref{S82footprint}). The SpIES footprint was chosen to take advantage of the SHELA footprint and for its relatively low background at mid-infrared wavelengths. As described in more detail in Section \ref{Complete}, background noise can drastically decrease the depth of the survey, which makes observing the faintest sources prohibitively difficult. 

SDSS observed S82 in five optical filters (\emph{ugriz}; \citealt{Fukugita1996}) to find variable objects and to obtain deeper imaging than the wider-area SDSS observations in the Northern Galactic Cap (\citealt{York2000}; \citealt{Frieman2008}; \citealt{Annis2014}). SDSS-I/II observed the full S82 field $\sim$80 times over 8 years resulting in photometry which reaches nearly two magnitudes fainter than the other fields in the survey (\citealt{Annis2014}, \citealt{Jiang2014}). S82 has also been observed multiple times with the SDSS spectrographs \citep{Smee2013} as part of the SDSS-I/II \citep{York2000} and SDSS-III/BOSS \citep{Eisenstein2011} campaigns, along with spectra from other facilities such as 2dF, 6dF, and AUS (\citealt{Croom2004, Croom2009}), WiggleZ \citep{Drinkwater2010}, the Virmos-VLT Deep Survey (VVDS; \citealt{LeFevre2005}), the VIMOS Public Extragalactic Redshift Survey (VIPERS1; \citealt{delaTorre2013}), DEEP2 \citep{Davis2007}, and the Prism Multi-Object Survey (PRIMUS; \citealt{Coil2011}). In total these facilities have collected $\sim$125,000 high quality spectra across its entire area. 

\begin{deluxetable*}{lcrcl}[!ht]
\tablecolumns{2}
\tablewidth{0pt}
\tablecaption{Deep imaging data available on Stripe 82}
\tablehead{
\colhead{Waveband}                                   & \colhead{Origin}  & \colhead{Depth} & \colhead{Coverage}      & \colhead{Reference} \\
\colhead{$\lambda_{\rm eff}$ ($\mu$m) }    & \colhead{}            & \colhead{}          & \colhead{  (deg$^{2}$)} & \colhead{} \\
}
\startdata
2-10 keV              & XMM-{\it Newton}            & 4.7$\times^{-15}$ erg s$^{-1}$ cm$^{-2}$  & 31.3$^{a}$ & \citet{LaMassa2015} \\
0.5-2 keV             & XMM-{\it Newton}            & 8.7$\times^{-16}$ erg s$^{-1}$ cm$^{-2}$  & 31.3$^{a}$  & \citet{LaMassa2015} \\
FUV, 1350--1750 \AA   & GALEX                       & $m_{\rm AB}\simeq23$                     & $\sim$200       & \citet{Martin2005}  \\
NUV, 1750--2750 \AA   & GALEX                       & $m_{\rm AB}\simeq23$                     & $\sim$200       & \citet{Martin2005}  \\
\vspace{4pt}
0.355 ($u$)           & SDSS                        & $m_{\rm AB}=23.90$                       & $\sim$300    & \citet{Jiang2014} \\
0.5 ($g$)             & SDSS                        & $m_{\rm AB}=25.10$                       & $\sim$300    & \citet{Jiang2014}  \\
                           &HSC$^{b}$                 &$m_{\rm AB}=26.50$                        &$\sim$300   &      Miyazaki et al.  \\
\vspace{4pt}
                           &DES                  &$m_{\rm AB}=26.50$                        &$\sim$300   & \citet{Diehl2014}  \\
0.6 ($r$)             & SDSS                       & $m_{\rm AB}=24,60$                 & $\sim$300    & \citet{Jiang2014}  \\
                           &HSC$^{b}$                  &$m_{\rm AB}=26.10$                        &$\sim$300   &  Miyazaki et al. \\
\vspace{4pt}
							          & DES                         & $m_{\rm AB}=26.00$                        & $\sim$300    & \citet{Diehl2014} \\
0.7 ($i$)             & SDSS    & $m_{\rm AB}=24.10$     & $\sim$300    & \citet{Jiang2014} \\
                           &HSC$^{b}$                  &$m_{\rm AB}=25.90$                        &$\sim$300   &     Miyazaki et al.   \\
								         & CS82			& $m_{\rm AB}=24.00$ 								& $\sim$170		& Kneib et al. in prep.\\
\vspace{4pt}
                          & DES    & $m_{\rm AB}=25.30$     & $\sim$300    &  \citet{Diehl2014} \\
0.9 ($z$)             & SDSS   & $m_{\rm AB}=22.80$                        & $\sim$300    & \citet{Jiang2014} \\
                           &HSC$^{b}$                  &$m_{\rm AB}=25.10$                        &$\sim$300   &  Miyazaki et al. \\
\vspace{4pt}
                           &  DES   &   $m_{\rm AB}=24.70$                         & $\sim$300    &\citet{Diehl2014}\\
1.00 ($Y$)            & ULAS$^{c}$               & $m_{\rm AB} = 20.93$            & 277.5          & \citet{Lawrence2007} \\ 
                           &HSC$^{b}$                  &$m_{\rm AB}=24.40$                        &$\sim$300   &     Miyazaki et al.    \\
                              &DES                    & $m_{\rm AB}=23.00$            &$\sim$300             &  \citet{Diehl2014}  \\ 
 \vspace{4pt}
                              &VHS                   &  $m_{\rm AB}=21.20$            & $\sim$300           & \citet{McMahon2013} \\ 
1.35 ($J$)            & ULAS$^{c}$                  & $m_{\rm AB} = 20.44$, 24 $\mu$Jy           & 277.5         & \citet{Lawrence2007} \\
									& VICS82,					&$m_{\rm AB} = 22.70$										& 150				&Geach et al. in prep.\\
\vspace{4pt}
                           &  VHS                        & $m_{\rm AB}=22.20$           & $\sim$300          &  \citet{McMahon2013} \\
1.65 ($H$)            & ULAS$^{c}$                       & $m_{\rm AB} = 19.98$, 37 $\mu$Jy           & 277.5         & \citet{Lawrence2007} \\
\vspace{4pt}
                            &  VHS                         &  $m_{\rm AB}=20.60$           &$\sim$300           &  \citet{McMahon2013} \\

2.20 ($K_{s}$)            & ULAS$^{c}$                       & $m_{\rm AB} = 20.10$, 33 $\mu$Jy           & 277.5      & \citet{Lawrence2007} \\
									& VICS82					&$m_{\rm AB} = 21.60$									& 150				&Geach et al. in prep.\\
\vspace{4pt}
                            & VHS                         & $m_{\rm AB}=21.50$           &  $\sim$300     &  \citet{McMahon2013} \\

3.6 (Ch1)         & {\bf SpIES}    & $m_{\rm AB}=21.90$ &  $\sim$115    & {\bf this paper} \\
\vspace{4pt}

                        & SHELA     &  $m_{\rm AB}=22.05$ &  $\sim$26    & \citet{Papovich2016} \\
4.5 (Ch2)         & {\bf SpIES}     & $m_{\rm AB}=22.00$ &    $\sim$115  & {\bf this paper}  \\
\vspace{4pt}

                        &  SHELA      & $m_{\rm AB}=22.05$  &     $\sim$26  &  \citet{Papovich2016} \\

250             & {\it Hershel}/SPIRE                   & 64.0, 64.0 mJy  & 270, 79     & \citet{Oliver2012, Viero2014}\\
350             & {\it Hershel}/SPIRE                   & 64.5, 64.5 mJy  & 270, 79     & \citet{Oliver2012, Viero2014}\\
500             & {\it Hershel}/SPIRE                   & 74.0, 74.0 mJy  & 270, 79     & \citet{Oliver2012, Viero2014} \\
1100 (277 GHz)            & ACT$^{d}$                                   &    $\sim$6.4 mJy    &       300       & analysis under way \\
1400 (218 GHz)           & ACT$^{d}$                                     &     $\sim$3.3 mJy      &         300    & \citet{Gralla2014, Das2014} \\
2000 (148 GHz)           & ACT$^{d}$                                     &     $\sim$2.2 mJy      &        300     & \citet{Gralla2014, Das2014} \\
21,000 (L-band) & VLA$^{e}$                                   & 260 $\mu$Jy     &     92        & \citet{Hodge2011}\\
30,000 (S-band)	& VLA$^{e}$                                   & 400 $\mu$Jy     &    $\sim$300      & \citet{Mooley2014} 
\label{Stripe82_data}
\enddata
\tablecomments{ $^{a}$Includes 7.4 deg$^{2}$ of archival \emph{Chandra} data,
$^{b}$Hyper Suprime-Cam (see \url{http://www.naoj.org/Projects/HSC/surveyplan.html} for more details),
$^{c}$UKIDSS Large Area Survey,
$^{d}$Atacama Cosmology Telescope,
$^{e}$Very Large Array
  }
\end{deluxetable*}

In addition to the collection of deep SDSS optical imaging (reaching a 5$\sigma$ AB magnitude of 24.6 in the $r$-band) and spectra, S82 contains a vast amount of multi-wavelength imaging taken over many epochs. The two panels of Figure \ref{S82footprint} show several multi-wavelength surveys that overlap with the SpIES region. At radio wavelengths, in addition to full coverage by the Faint Images of the Radio Sky at Twenty-centimeters (FIRST; \citealt{Becker1995}, \citealt{Helfand2015}) survey,  \citet{Hodge2011} provided 1$\farcs$8 resolution data down to 52$\mu$Jy at 1.4GHz (L-band) over $\sim$90 deg$^2$ of Stripe 82 (twice the resolution and three times the depth of FIRST).  Additional radio data will be forthcoming at lower resolution (e.g.,\ \citealt{Jarvis2014}) and at higher frequency \citep{Mooley2014}. 

In the far-infrared, the {\emph{Herschel Space Observatory}} performed the HerMES Large Mode Survey (HeLMS; \citealt{Oliver2012}) and the Herschel Stripe 82 Survey (HerS; \citealt{Viero2014}) to study galaxy formation and correlations between galaxies and dark matter haloes. Existing mid-infrared observations of S82 include SHELA \citep{Papovich2016}, which contains deep imaging data for dark energy measurements, and the AllWISE observations from \emph{WISE} \citep{Wright2010}. Near-infrared measurements of S82 have been performed by the UKIRT Infrared Deep Sky Survey (UKIDSS; \citealt{Lawrence2007}), the VISTA Hemisphere Survey (VHS; \citealt{McMahon2013})---which is matched to the SDSS coadd photometry in the catalog presented in \citet{Bundy2015}---and the deeper J- and K-band coverage from the VISTA-CFHT Stripe 82 Survey over 130 deg$^{2}$ of S82 (VICS82; Geach et al.\,in prep.). In addition to SDSS, Stripe 82 has high-resolution imaging (median seeing of 0$\farcs$6) from the CFHT Stripe 82 Survey (CS82; Kneib et al.\,in prep.) and is part of the Dark Energy Survey\footnote{\href{DES}{http://www.darkenergysurvey.org/}} (DES) footprint. 

S82 was also mapped in the ultraviolet as part of the {\emph{GALEX}} All-sky Imaging Survey and Medium Imaging Survey, and a few locations were imaged with the Deep Imaging Survey as outlined in \citet{Martin2005}. \emph{Chandra} and \emph{XMM-Newton} have been used to observe partly contiguous regions over a wide area at X-ray wavelengths, searching for high luminosity quasars (\citealt{LaMassa2013b, LaMassa2013c}), with the most recent large-area X-ray catalog release covering $\sim$31\sqdeg with \emph{XMM-Newton} \citep{LaMassa2015}. More observations are cited in Table \ref{Stripe82_data} which lists some properties of the deepest imaging data of S82 at various wavelengths.  The combination of all of the multi-epoch, multi-wavelength spectroscopic and photometric data on S82 provides a powerful tool to aid in our understanding of the Universe by painting a multi-wavelength and multi-epoch picture of matched objects between these surveys.


\section{Data Acquisition}\label{sec:Obs_Strat}

SpIES data were obtained as part of Cycle 9 (2012-2014) of the \Spitzer `warm' post-cryogenic mission utilizing the first two channels of IRAC. IRAC is a wide-field camera with four channels, each 256$\times$256 pixels with a 5$\farcm$2$\times$5$\farcm$2 field of view \citep{Fazio2004b}. The first two arrays (3.6 and 4.5 microns) are designed to observe the sky simultaneously, which decreases observation time and ensures that the epochs of measurement are roughly the same for both channels. \Spitzer has been operating in `warm' mode long enough to measure and report the differences in IRAC performance between the cryogenic and `warm' observations\footnote{\href{WarmSpitzer}{http://irsa.ipac.caltech.edu/data/SPITZER/\\docs/irac/warmimgcharacteristics/}}. The changes in performance, including changes in PSF, sensitivity levels, and constant values such as gain and flux conversion, are minor and the overall performance of IRAC has not degraded substantially with time (see \citealt{Mauduit2012}). 

\begin{figure*}[ht]
\centering
\includegraphics[scale=0.4]{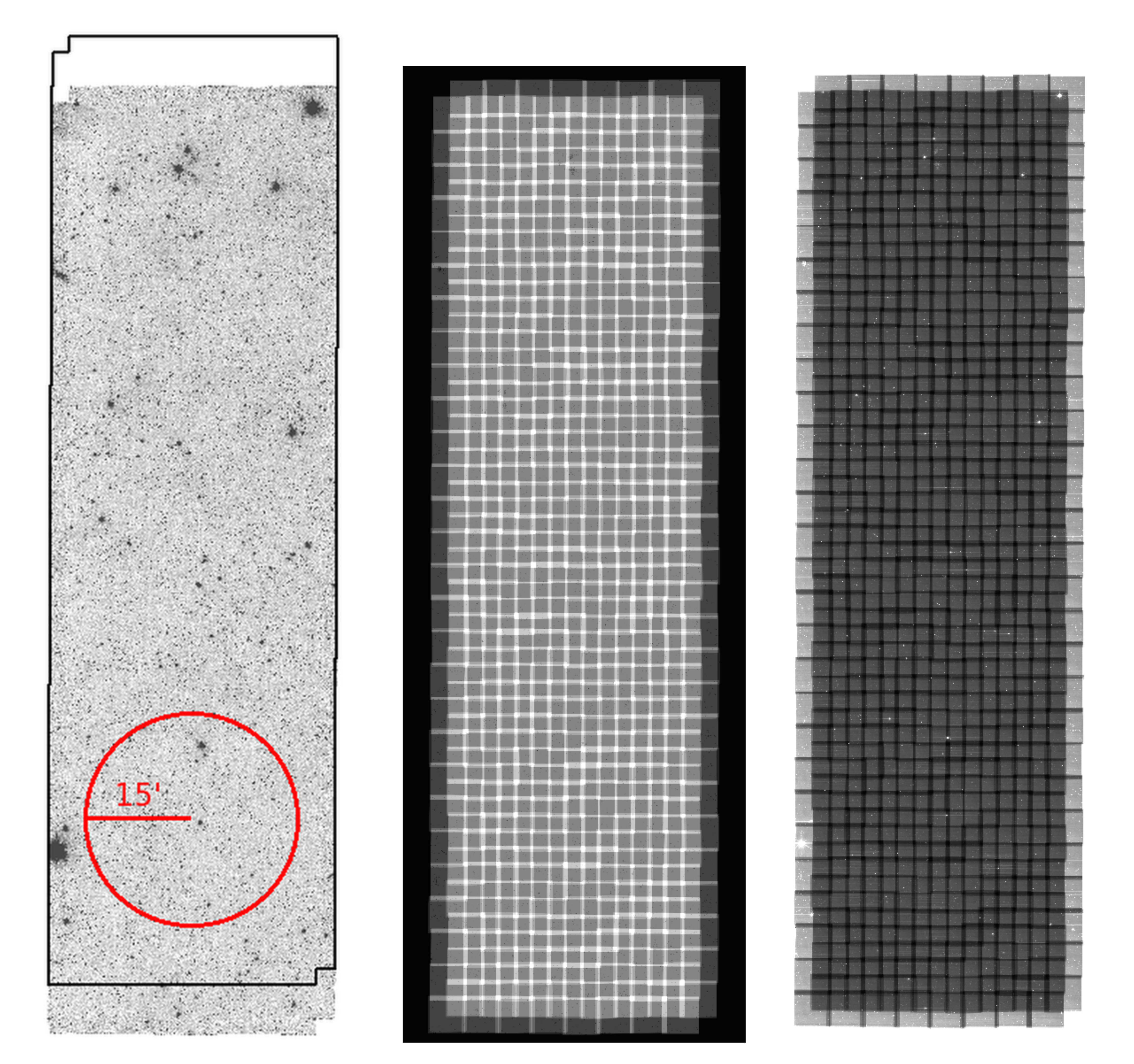}
\caption{\footnotesize{Left: One SpIES 3.6\,$\mu$m, double-epoch, stacked AOR from which we extract sources. This is one of 77 stacked AORs (154 single epoch AORs divided by two epochs) that are strung together (see Figure \ref{S82footprint}) to cover the entire SpIES field. The red circular region illustrates the angular size of the Moon, and the black region shows the coverage of the same AOR at 4.5\,$\mu$m. Center: An example of the coverage map of the AOR, showing where the individual pointings of IRAC overlap when they are combined to form the AOR. These maps are unique to each AOR and are used as weighted images during source extraction. Pixels with lighter colors have more coverages. The AOR footprint has been padded with a band corresponding to zero coverage. Right: The flux density uncertainty map of each AOR, where the values only take into account details in pipeline processing error propagation, not source extraction. In this map, darker colors correspond to lower uncertainties in flux density. The lower uncertainties align with the higher coverage values shown in the central panel.}}
\label{AORpic}
\end{figure*}

\begin{deluxetable}{lr}[h] 
\tablecolumns{2}
\tablewidth{0pt}
\tablecaption{Astronomical Observation Request (AOR) Time Table \label{time}}
\tablehead{
\colhead{Operation} & \colhead{Time (s)}}
\startdata
Exposure time at each pointing  & 30          \\ 
$\times$2 dithering            & 60          \\ 
$\times \sim$224 pointings          & 13440         \\ 
$+$ Slew Time                   & $\sim$2400        \\ 
$+$ Settle Time                 & $\sim$2400        \\ 
$+$ Overhead(Slew and Download) & $\sim$600      \\ 
\hspace{1em}$\times$2 epochs               & $\sim$37700     \\ 
\hspace{1em}$\times$77 AORs                & $\sim$2.9$\times10^{6}$ \\ 
& \\
Total Observation Time          & $\sim$820hr          
\enddata
\tablecomments{Approximate exposure time breakdown for SpIES for each detector (the larger AORs required more time than estimated). The two dithers and the two epochs combined with 30s exposures each lead to a total AOR exposure time of $2\times2\times30=120$s for both channels. SpIES spent $\sim$70$\%$ of the time in observation and $\sim$30$\%$ in motion to other fields.}
\end{deluxetable}

The SpIES observation strategy was motivated by the strategies of previous \Spitzer campaigns such as SDWFS \citep{Ashby2009}, SWIRE \citep{Lonsdale2003}, SERVS \citep{Mauduit2012}, and SSDF \citep{Ashby2013}. Similar to these surveys, SpIES observations were separated into individual Astronomical Observation Requests (AORs), which are self-contained exposure sequences executed independently of each other. AORs are comprised of sequential pointings of IRAC which are stacked to form a single image. AORs overlap slightly, to form the entire field (see the SpIES regions in Figure \ref{S82footprint}). Most of the SpIES AORs consist of a map of 8$\times$28 IRAC FOVs, corresponding to a total area of $\sim$1.63 \sqdeg per AOR (see Figure \ref{AORpic}). There were, however, a few AORs which needed to be adjusted in width due to changes in position angle between AOR observations (observations separated by $\sim$6 months have a field rotation of $\sim$180$\deg$), to connect with their neighboring AORs and form a continuous strip. Four of our AORs were increased to 9$\times$28 pointings, two were increased to 10$\times$28 pointings, and one was decreased to 5$\times$28 pointings. The size differences can be identified by an increase or decrease of the given AOR integration time in Appendix \ref{AppendB}. In total, SpIES is comprised of 154 AORs observed over two epochs (77 AORs per epoch) which corresponds to $\sim$70,000 IRAC FOVs spanning the full survey area.

Each AOR was built by successively pointing and dithering IRAC until the 8$\times$28 map was complete, using a small-cycle dither pattern. This pattern offsets the observations by up to 11 pixels ($\sim$13$\arcsec$) to obtain overlapping coverage while eliminating some instrumental problems such as bad pixel detections and bright star saturation \citep{Mauduit2012}. Built into the cycle dither pattern is a sub-pixel dither pattern of half a pixel, which improves the 1$\farcs$2 per pixel sampling to 0$\farcs$6 per pixel after the images are stacked. This oversampling reduces effects that bad pixels and bright star saturation have on the image. This issue must be accounted for when calculating source flux error in Section \ref{sec:errors}. 

Images are taken simultaneously at 3.6\,$\mu$m and 4.5\,$\mu$m with a $\sim$6$\farcm$7 offset between the two channels due to the physical placement of the arrays. This offset leads to a section around the perimeter where objects are detected in one band and not the other (as shown in Figure \ref{AORpic}). The catalogs described in Section \ref{fullcat} indicate which objects lack a counterpart in the other band due to these regions without overlapping dual-band coverage. Additionally, the survey area  changes slightly due to this offset. The quoted area of $\sim$115 deg$^{2}$ is the coverage where SpIES detects sources at either 3.6\,$\mu$m or 4.5\,$\mu$m. The coverage of each individual detector is $\sim$107 deg$^{2}$ where the coverage of the overlap of the two detectors (detections at both 3.6\,$\mu$m and 4.5\,$\mu$m) is $\sim$100 deg$^{2}$. This is important when computing number densities in Section \ref{Complete}.

\begin{figure}[h!]
\centering
\includegraphics[width=7cm, height=7cm, trim=5mm 5mm 5mm 5mm,]{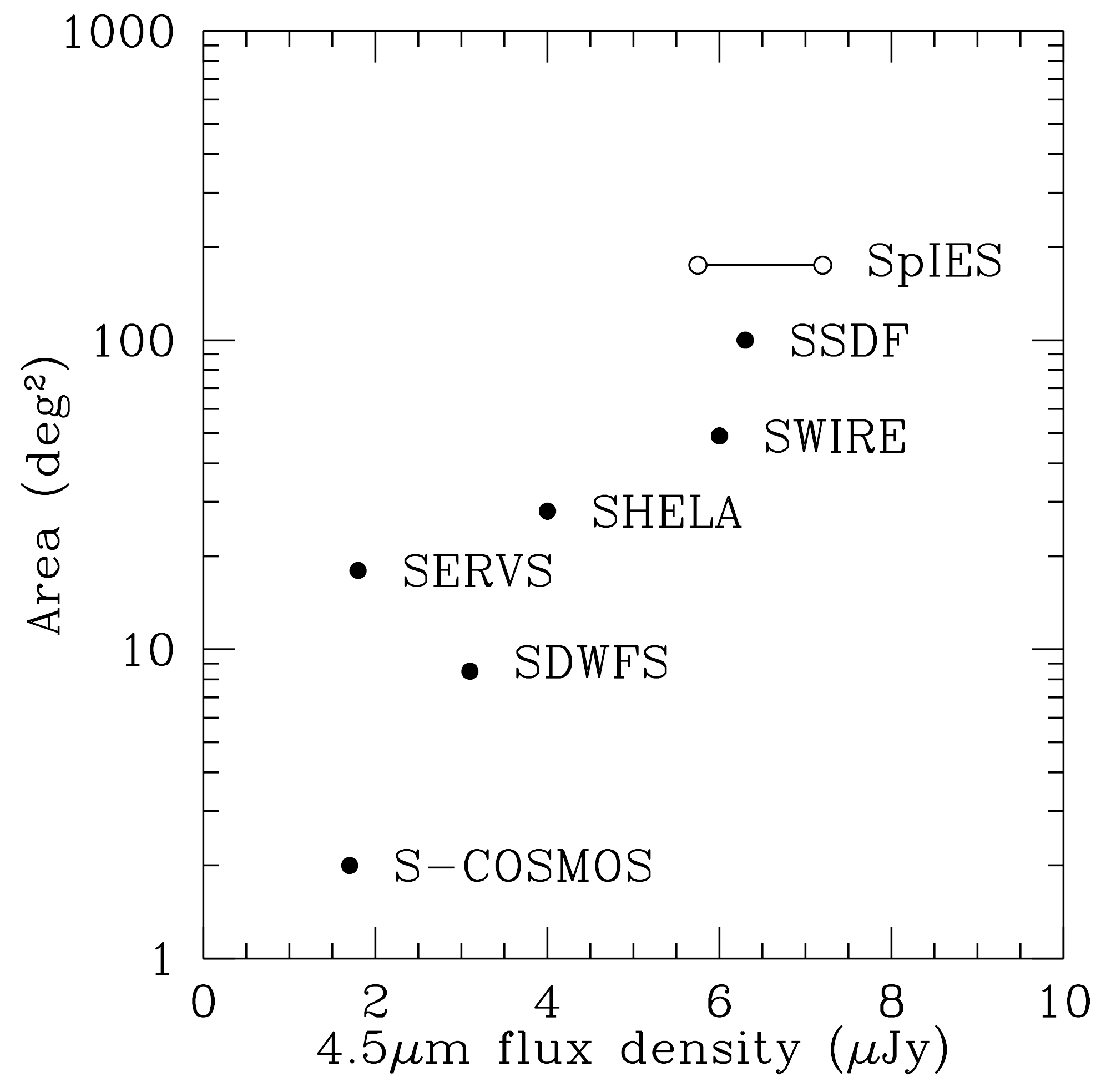}
\caption{\footnotesize{Comparison of the calculated 4.5 $\mu$m 5$\sigma$ depth to area of the major \Spitzer surveys. Depths are calculated using the \Spitzer Sensitivity Performance Estimation Tool (SENS-PET) assuming a low background. At $\sim$115 \sqdeg in area SpIES is the largest \Spitzer survey and probes SWIRE depths \citep{Lonsdale2003}. Open circles show the measured depth (left; see Table \ref{Complevel}) and calculated depth from SENS-PET with a medium background (right) for SpIES.}}
\label{depth_area}
\end{figure}

Observations were performed over two distinct epochs separated by no less than five hours in time (see Appendix \ref{AppendB}) and shifted by half a FOV in both right ascension and declination. Multiple epoch observations allow for detection of transient objects, and the spatial offset ensures that detected objects are observed on different regions of the array, allowing for more accurate photometry. In most cases, the second epoch of observation was taken directly after the first, where the observation time for the first epoch of a full AOR ($\sim$5 hours including slew and settle time) was sufficient to significantly separate the two epochs. For a typical asteroid, which moves at \mbox{$\sim$25$\arcsec$ hr$^{-1}$ \citep{Ashby2009}}, a five-hour temporal separation leads to $\sim$2$\arcmin$ spatial separation, which is easily detected in separate epochs. The SpIES field is covered with at least four exposures at each pixel, providing both deep and reliable photometry across the large area of observation---with an exception around the perimeter where the second epoch has been shifted by half a FOV.

The SpIES AORs were constructed to maximize area while maintaining a depth comparable to that of SWIRE \citep{Lonsdale2003}. To achieve this goal, each AOR was observed for a total of 60 seconds, split evenly among the two dithered pointings of 30 seconds each. The limiting flux does not reach the IRAC confusion limit, and therefore confusion noise, which does not decrease as the square root of exposure time \citep{Surace2005}, is small (see Section \ref{Conf} for more detail). The total observation time for the SpIES survey was $\sim$820 hours (Table \ref{time}) split among the 154 AORs. Figure \ref{depth_area} demonstrates that the SpIES survey is both the largest \Spitzer survey to date and reaches approximately to SWIRE depths, fulfilling two of the projects primary goals.

\section{Image Reprocessing}\label{sec:DatRed}
\begin{figure}[t!]
\centering
\includegraphics[scale =0.12]{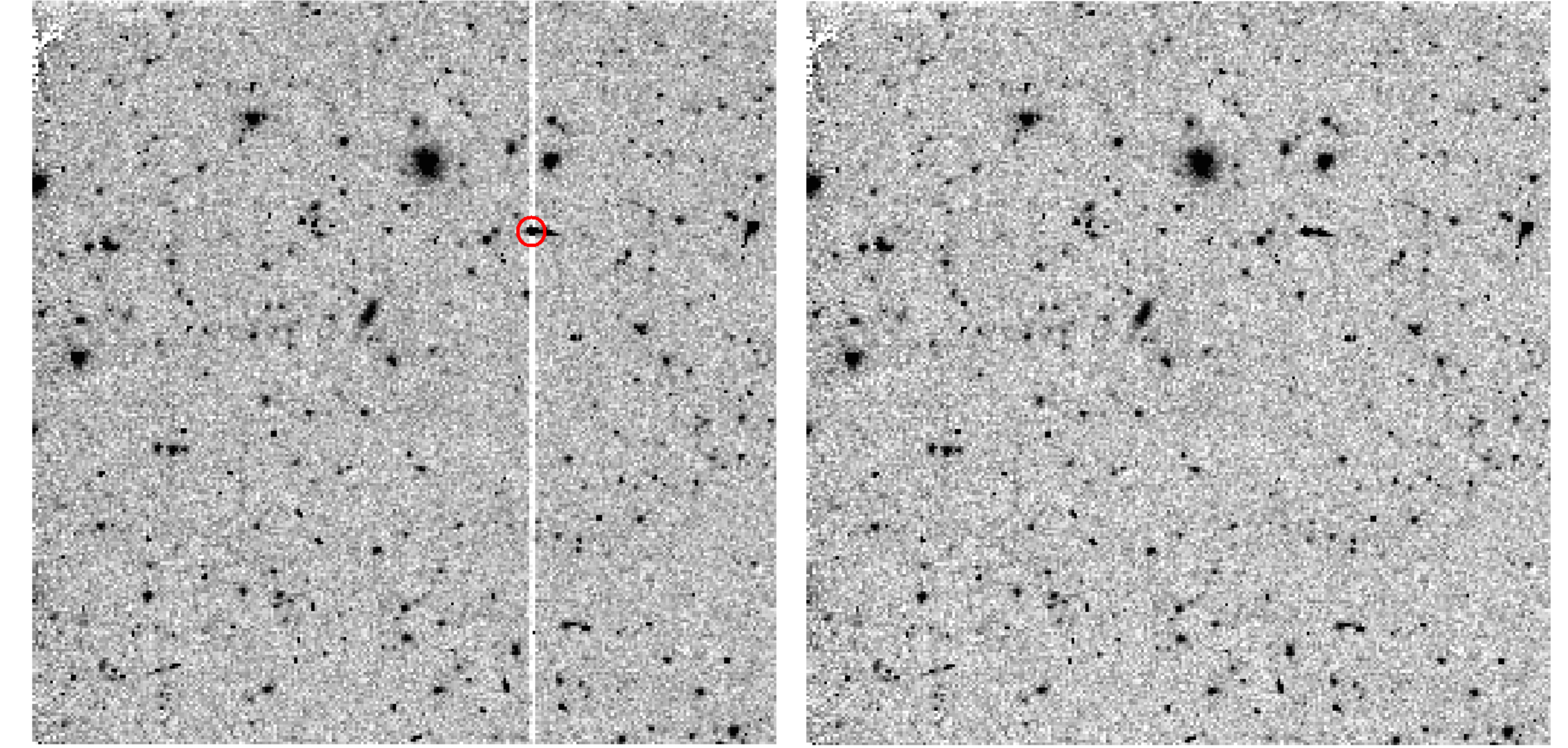}
\caption{\footnotesize{Left: Typical SpIES Level 1 BCD image from the SSC before corrections. The bright pixel (red circle) causes its whole column to drop to a low background value (causing the white line across the full array). Right: A cBCD image, which is the BCD image after it has been corrected for known signatures, such as the column pulldown in the left panel. The cBCD images are the size of an IRAC FOV (5$\farcm$2$\times$5$\farcm$2) and are mosaicked together to form the larger AORs seen in Figure \ref{AORpic}. Both images are centered at \mbox{($\alpha$, $\delta$)=(32.611, -0.887)} degrees.}}
\label{BCD}
\end{figure}

Observations from \Spitzer are downlinked to the \Spitzer Science Center (SSC) where the raw images are sent through the ``Level 1" processing pipeline. This pipeline corrects for known instrumental signatures in the images (dark subtraction, ghosting, and flatfielding) and flags possible cosmic ray hits. Additionally, the observed counts units (ADU) are converted into flux density units (MJy sr$^{-1}$), creating the Basic Calibrated Data (BCD) images (see Section 5 of the IRAC Handbook\footref{IRAC}). These BCD images are processed one 5$\farcm$2$\times$5$\farcm$2 field at a time through a secondary pipeline to correct for other artifacts seen in IRAC images such as stray light (masking of scattered light from stars outside the array location) and column pulldown (a bright pixel causing a low background in the CCD array column; Figure \ref{BCD}). The resulting Corrected-BCD (cBCD) images (Section 6 of the IRAC Handbook) were used to create stacked AORs in SpIES (see Figure \ref{AORpic}). A single cBCD image only covers one IRAC FOV; however, after accounting for the dithers and the two epochs, we have a total of four cBCD images which cover roughly the same region of the sky. The cBCD images are stacked to create the larger AOR mosaics using the SSC Mosaicing and Point-source Extraction (MOPEX\footnote{\href{MOPEX}{http://irsa.ipac.caltech.edu/data/SPITZER/docs/\\dataanalysistools/tools/mopex/mopexusersguide/}}) software. 

\begin{deluxetable}{llr}
\tablecolumns{3}
\tablewidth{0pt}
\tablecaption{Parameter values for Mopex and SExtractor \label{Params}}
\tablehead{
\colhead{Program} & \colhead{Parameter} & \colhead{Value}}
\startdata
MOPEX & Fatal\_Bitpattern & 27392$^a$ \\
SExtractor & DETECT\_THRESH & 1.25\\
SExtractor & DETECT\_MINAREA & 4 \\
SExtractor & DEBLEND\_NTHRESH & 64 \\
SExtractor & DEBLEND\_MINCONT & 0.005 \\
SExtractor & PHOT\_APERTURES$^{b}$ & 4.8, 6.4, 9.63,\\&& 13.6, 19.2, 40 \\
SExtractor & PIXEL\_SCALE & 0.6 \\
SExtractor & BACK\_SIZE & 64 \\
SExtractor & BACK\_FILTERSIZE & 5  \\
SExtractor & GAIN & 4429.37, 3788.29$^{c}$ \\
SExtractor & WEIGHT\_TYPE & MAP\_WEIGHT\\
SExtractor & WEIGHT\_IMAGE & mosaic\_cov.fits \\
SExtractor & WEIGHT\_GAIN & Y \\
SExtractor & FILTER & Y \\
SExtractor & FILTER\_NAME & default.conv
\enddata
\tablecomments{Parameters that were changed from the default MOPEX or SExtractor configuration files. These parameters were used in the stacking and source extraction of the SpIES images.\\$^{a}$DCE\_Status\_Mask\_Fatal\_BitPattern with bits 8,9,11,13,14 are turned on.\\$^{b}$ The diameter of the aperture in pixels. \\$^{c}$Gain values for the 3.6\,$\mu$m, 4.5\,$\mu$m detector. See Section \ref{sec:errors} for more details}
\end{deluxetable}

The MOPEX software was developed by the SSC specifically to process \Spitzer BCD and cBCD images. This package contains several pipelines which can be used to process, stack, and extract sources from \Spitzer images; however, we only relied on the mosaic pipeline to combine cBCD images onto a common frame. There are five stages of combination in the mosaic pipeline which transform a list of cBCD images to a full mosaic. First, an interpolation technique is run on the input images, determining the location of each pixel and forming a fiducial frame for the output image. Next, an outlier rejection script is run which flags or masks bad pixels from the final image. These flags are applied to the fiducial frame with a re-interpolation technique. Co-addition of pixel values is performed on tiles of pixels that make up the full image using a method defined by the user (for SpIES, pixels were co-added using a straight average). Finally, a script combines the tiles from the co-addition stage together to form a single image. Along with a combined image, MOPEX provides an option to output other datasets such as a coverage map and uncertainty map similar to those shown in Figure \ref{AORpic}. The SSC also provides these images as ``Level 2" post-BCD (pBCD) images which have been processed by MOPEX and thus can be used for source extraction and photometry; however, they are only single epoch images, thus do not achieve the full depth of our survey.

\begin{figure}[t!]
\centering
\includegraphics[scale =0.15]{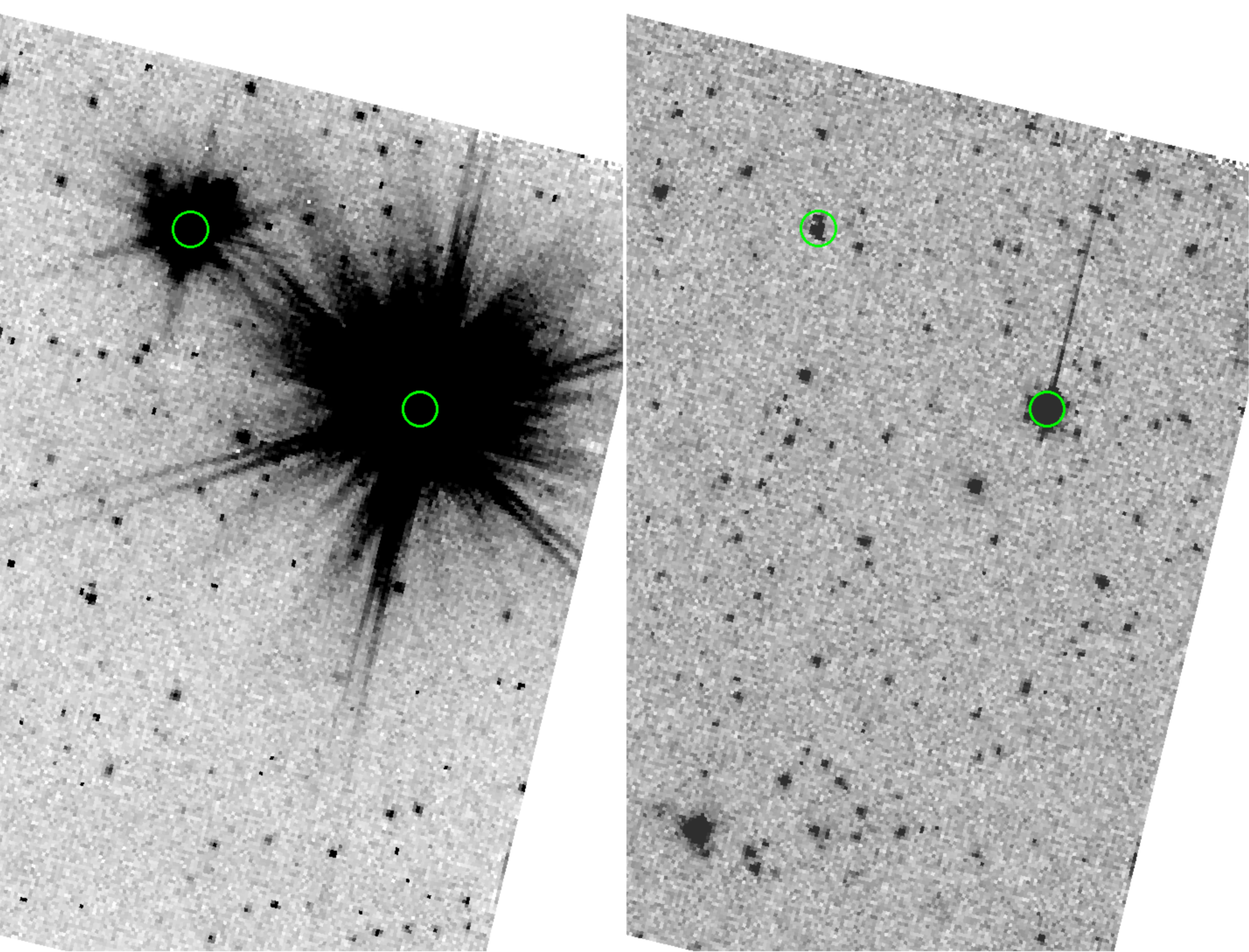}
\caption{\footnotesize{Shown on the left is an example of two bright stars in a $\sim$3$\arcmin \times$4$\farcm$5 cutout of a 3.6\,$\mu$m cBCD (centered at \mbox{($\alpha$, $\delta$)=(34.464, -0.169)} degrees). The image in the right panel is the next observation (centered at \mbox{($\alpha$, $\delta$)=(34.482, -0.247)} degrees) showing the latent images from the bright stars in the previous observation (left panel). The green circles highlight the pixel location of the latent objects in IRAC from subsequent observations at different sky locations.}}
\label{latentBCD}
\end{figure}

\begin{figure}[t!]
\centering
\includegraphics[scale =0.15]{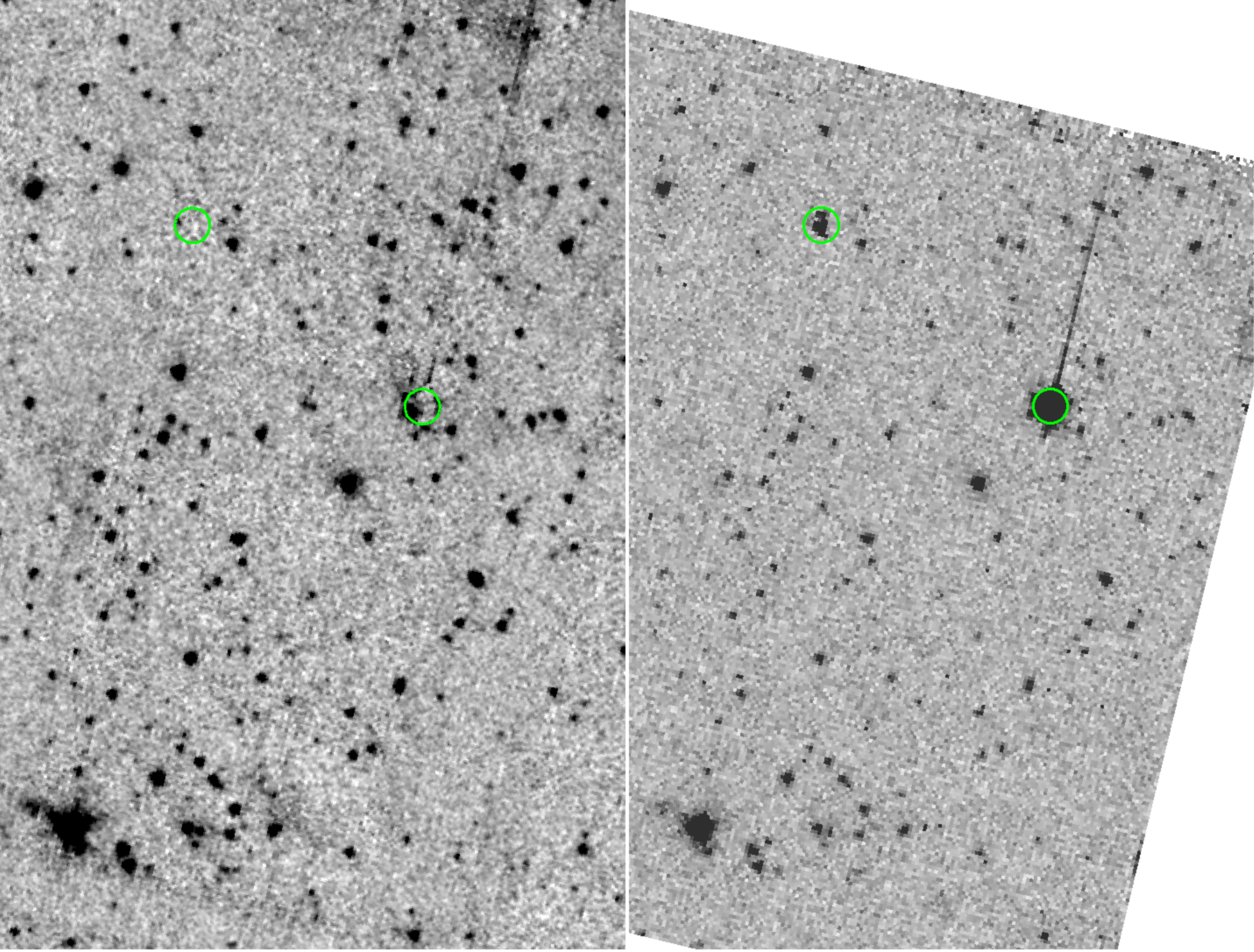}
\caption{\footnotesize{Here, the left panel shows a portion of the final stacked AOR image after sky matching to the right panel in Figure \ref{latentBCD} (also the right panel of this figure) with the latent object locations outlined in green. The latent objects in the cBCD (right panel) are masked in the final stacked image (left panel) because the latent image bits were turned off in the MOPEX processing pipeline (see Table \ref{Params}), therefore, they do not appear in the final catalogs.}}
\label{latentmask}
\end{figure}

To achieve our full depth, we created images by submitting the cBCD images of the two overlapping epochs as well as their corresponding bit mask (bimsk) images and the uncertainty (cbunc) into MOPEX. The pipeline was run using the default parameters with the exception of the DCE\_Status\_Mask\_Fatal\_BitPattern (see Table \ref{Params}) which tells MOPEX which pixels to mask in the final mosaic based on the bit value of those pixels in the input bit mask. For example, the 3.6\,$\mu$m `warm' IRAC images suffer from latent images\footnote{\href{warmspitz}{http://irsa.ipac.caltech.edu/data/SPITZER/\\docs/irac/iracinstrumenthandbook/63/}} (typically after exposure to bright stars) which remain at the same pixel location on the detector for the next set of observations (see Figure \ref{latentBCD}). If left unchecked, these objects appear in a different sky location in the final image, and will be detected as individual sources. To prevent contamination in the final AOR, the SSC pipeline locates latent objects in each BCD, and flags the corresponding pixels in the bit mask\footnote{\href{IRACHB}{http://irsa.ipac.caltech.edu/data/SPITZER/\\docs/irac/iracinstrumenthandbook/44/\#\_Toc410728355}} for that BCD. We then set the DCE\_Status\_Mask\_Fatal\_BitPattern (which reads the bit masks) to mask any objects that have that particular flag set in the final combined image (see Figure \ref{latentmask}). Since latent objects do not appear in our final stacked images they are not present in our final catalogs.

\begin{figure}[h]
\begin{center}
  \includegraphics[scale=0.15]{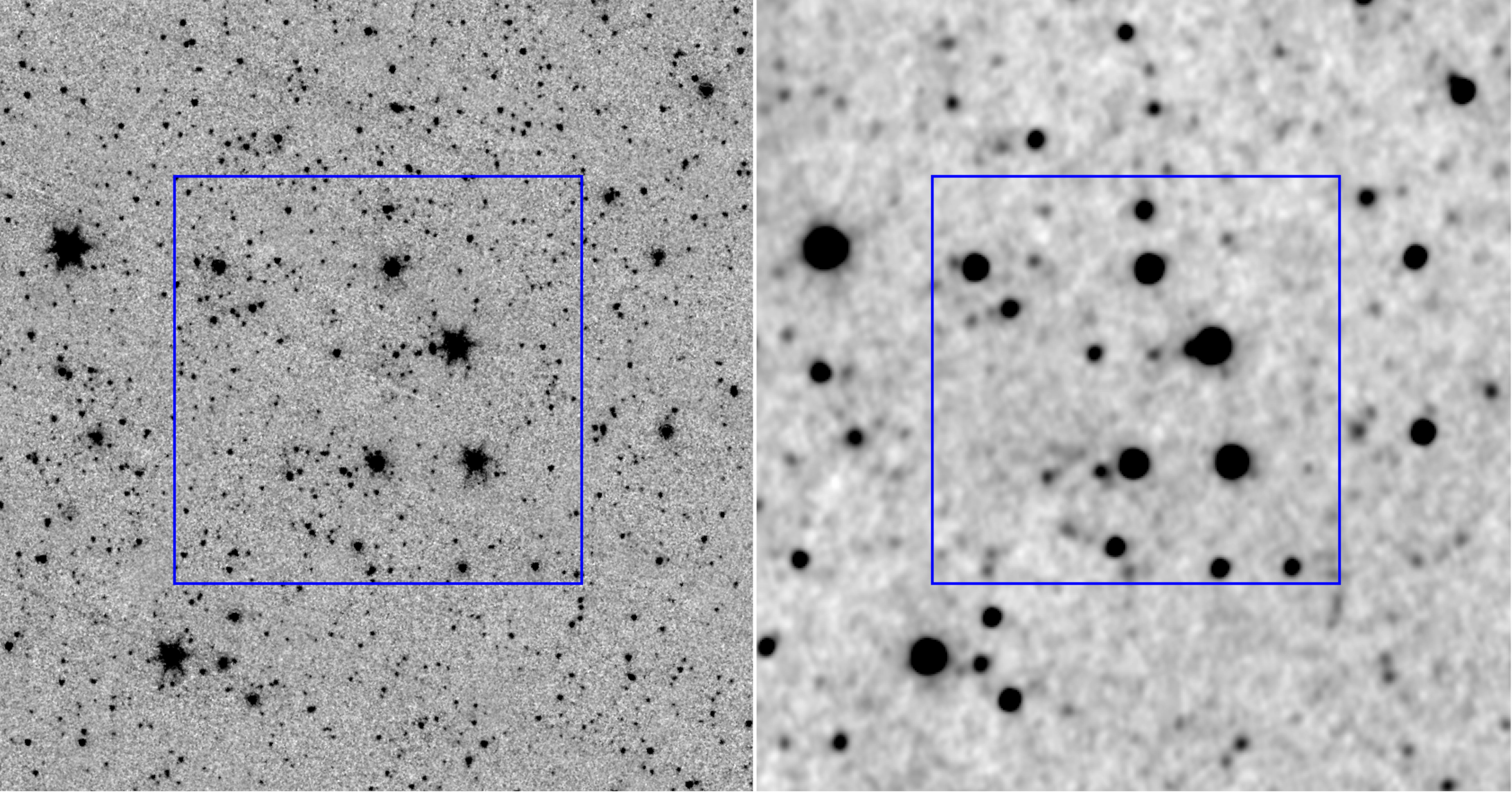}
\caption{Comparison of a $\sim$100 arcmin$^{2}$ box of a SpIES 4.5\,$\mu$m image and a 4.6\,$\mu$m image which cover approximately the same central wavelength. `Warm' IRAC 4.5\,$\mu$m has a PSF of 2$\farcs$02 compared to 6$\farcs$4 for \emph{WISE} 4.6\,$\mu$m, allowing SpIES to resolve objects that are blended in \emph{WISE}. Additionally, the superior depth of SpIES (AB magnitude of $\sim$22 in [4.5] compared to $\sim$18.8 in W2) yields more sources above the background ($\sim$1400 in the dual-band catalog) in the field shown compared to \emph{WISE} ($\sim$350 in AllWISE). The blue boxes represent a single FOV of IRAC ($5\farcm2 \times 5\farcm2$). }
\label{fig:spieswise}
\end{center}
\end{figure}

The SSC-produced BCD, cBCD, and pBCD images, as well as all ancillary data images (uncertainty maps, coverage maps, etc.), are publicly available on the \Spitzer Heritage Archive\footnote{\href{SHA}{http://sha.ipac.caltech.edu/applications/Spitzer/SHA/}} (SHA) website. The images created by the SpIES team are publicly available (see Appendix \ref{AppendA}). There are a total of 231 images created by the SpIES team consisting of 154 individual epoch AOR mosaics and 77 combined epoch mosaics (stacking the two overlapping individual epoch images). Source extraction and photometry were performed on each of these 231 images. The final catalogs were constructed by running our source extraction techniques on the 77 combined epoch AORs to take advantage of the full depth of SpIES. To illustrate the depth of SpIES, Figure \ref{fig:spieswise} compares a region from a full-depth 4.5 micron AOR and the same region from \emph{WISE} 4.6 micron ($W2$).

\section{Catalog Production}\label{catalog}
\subsection{Source Extraction}\label{Extraction}
The SpIES catalogs were constructed by running Source Extractor (SExtractor; \citealt{BertinSE}) on each combined-epoch AOR mosaic, creating 77 AOR source catalogs for the 3.6\,$\mu$m detections and 77 for the 4.5\,$\mu$m detections. SExtractor uses a six-step source extraction routine which efficiently generates catalogs from large images. First, a robust 3$\sigma$ clipped background estimation is performed on the entire image, which has been inspected through an output background map. This step is followed by a thresholding algorithm which extracts objects at a certain, user-specified standard deviation above the background. SExtractor then runs a deblending routine to separate potentially blended sources, filters the image using an input filtering routine, and performs photometry on detected sources within user specified apertures. Finally, SExtractor attempts to classify objects as point-like (stars) or extended (galaxies) based on the input pixel scale and stellar FWHM of the survey. 

Each step is controlled through an input configuration file and an output parameter file. There are a variety of parameters that can be changed in the configuration file, some of which can significantly change the source extraction results. The final configuration file was a mix of parameters extensively tested on the SpIES images and parameters adopted from previous programs such as the SERVS \citep{Mauduit2012} and SWIRE \citep{Lonsdale2003} surveys. Table \ref{Params} lists the configuration parameters used in our processing.

Previous \Spitzer surveys also used the coverage map created in MOPEX as a weighted image during source extraction. These images hold information about the number of times a particular pixel in the AOR was observed, which is related to the effective exposure time at each pixel. Since the signal-to-noise ratio of an object increases with the square root of exposure time in these data, the coverage maps assign pixels with more coverages (i.e., longer exposures) a higher weight. Following this convention, the coverage maps were input as weight maps, converted into a variance map by SExtractor through the inverse relationship between weight and variance, and scaled to an absolute variance map created internally by SExtactor. This processing is also controlled through the input configuration file during source extraction. 

SExtractor can be run in either single-detection mode, which performs source detection, aperture definition, and photometry on the same image, or dual-detection mode, which finds sources and defines apertures on a first input image (for example, a 3.6\,$\mu$m AOR) and performs photometry on a second input image (the same AOR observed using the 4.5\,$\mu$m detector). All of the SpIES AOR mosaics were run in single-detection mode, creating 77 double-epoch catalogs for each channel. Full-area, single-channel catalogs were made by concatenating the 77 individual AOR catalogs using the Starlink Tables Infrastructure Library Tool Set (STILTS)\footnote{\href{stilts}{http://www.star.bris.ac.uk/$\sim$mbt/stilts/}}. These single-channel catalogs are designed to contain a single row for each object in the SpIES survey, so when two objects match within 1$\arcsec$ between two AORs (which is possible since the AORs overlap) we report the average position, the weighted average of the flux density values (using the errors as weights), and the errors added in quadrature in a single row in the catalog (the overlapping regions between AORs account for $\sim$10\% of the total survey area). Though we report objects that are detected 5$\sigma$ above the calculated background, many objects have a signal-to-noise (S/N) less than 5 due to Poisson noise.

\begin{deluxetable}{cccccc}[t]
\tablecolumns{6}
\tablewidth{0pt}
\tablecaption{Aperture correction for SpIES}
\tablehead{
\colhead{Band} & \colhead{$1\farcs4$} & \colhead{$1\farcs9$} & \colhead{$2\farcs9$} & \colhead{$4\farcs1$} & \colhead{$5\farcs8$}}
\startdata
$3.6\,\mu$m&0.584&0.732&0.864&0.911&0.950 \\
$4.5\,\mu$m&0.570&0.713&0.865&0.906&0.946
\enddata
\tablecomments{Measured aperture corrections for SpIES objects with good flags matched to the 2MASS point source catalog. These corrections are nearly identical to those used in SERVS \citep{Mauduit2012} for identical aperture radii.}
\label{tab:aper}
\end{deluxetable}

Photometry on SpIES sources was performed in six circular apertures of radii $1\farcs4$, $1\farcs9$, $2\farcs9$, $4\farcs1$, $5\farcs8$, and $12\arcsec$, reported as diameter in pixels in the SExtractor configuration file in Table \ref{Params}. The first five apertures (which are the same size as the SERVS apertures) contain only a fraction of the light from each source, while the sixth contains ``all" the light from the source (see Section 4.11 of the IRAC Handbook\footref{IRAC}). The aperture correction factors in Table \ref{tab:aper} are measured for the SpIES survey for objects with good flags (discussed in more detail in Section \ref{fullcat}) matched to the 2MASS Point Source Catalog (PSC) to ensure that measurements were performed on point sources only. We then took the ratio of the light in the smaller apertures to the light in the largest aperture, made a histogram of the resulting factors for each aperture, and fit a Gaussian to that histogram to measure the peak and spread of the distribution. The location of the peak of the Gaussian was used as the correction factor. The corrections measured for SpIES differ by less than 1\% of those used in SERVS \citep{Mauduit2012} for the exact same aperture radii. Aperture corrections are useful for finding faint objects with a radius much less than the large 12$\arcsec$ radius aperture, because in these cases the background noise in the aperture would dominate the object. We primarily use the 1$\farcs$9 radius aperture for analysis in the following sections as it corresponds to a $\sim$70\% curve of growth correction (the curve showing how the flux density ratio changes with aperture size) in both channels. 

After objects are extracted from the images, the surface brightness values are converted from the \Spitzer image unit of $\mathrm{MJy\,sr}^{-1}$ to flux densities ($\mu$Jy) per pixel using the following conversion:  

\begin{equation}\notag
\mathrm{\frac{MJy}{sr}}\left(10^{12}\frac{\mu \mathrm{Jy}}{\mathrm{MJy}}\right)\left(\frac{\pi \mathrm{rad}}{180\deg}\right)^2\left(\frac{1\deg}{3600\arcsec}\right)^2\left(\frac{0\farcs6}{\mathrm{pixel}}\right)^2
\end{equation}
such that,
\begin{equation}
1\mathrm{MJy \,steradian}^{-1}=8.46\mu\mathrm{Jy}\,\mathrm{pixel}^{-2}
\label{eq:convert}
\end{equation}
where we multiply by the SpIES pixel size of 0$\farcs$6, which is half of the IRAC pixel size due to the image dithering.

This correction factor in Equation \ref{eq:convert} was applied to each pixel in the image which, when summed in an aperture, yields the total flux density of the source. This value was divided by the appropriate aperture correction from Table \ref{tab:aper} to produce the final flux density value for the objects in the catalogs.

\subsection{Photometric Errors}\label{sec:errors}
Photometric errors were computed using SExtractor and are reported in the catalog (see Table \ref{Params}). According to Section 10.4 of the SExtractor manual, the $1\sigma$ photometric errors are computed via
\begin{equation}\label{SEerr}
\sigma _{\rm{source}}= \sqrt{A\sigma_{rms} ^2 + \frac{F}{g}},
\end{equation} 
where $A$ is the measurement area in pixels, $\sigma_{rms}$ is the background root-mean-square (rms) value of each pixel, $F$ is the background-subtracted source count value in the measurement aperture, and $g$ is the detector gain.  This expression is simply the rms background added in quadrature with the Poisson noise. SExtractor assumes that the signal in the input images is in units of counts, typically a Digital Number (DN) which is the number of photons counted scaled by the detector gain value. \Spitzer images, however, are converted to physical units during ``Level 1" processing. Many previous surveys which have used SExtractor to compute photometric errors exclude the Poisson noise and only report the rms background error, which is also the SExtractor default if no gain is supplied. For bright objects, Poisson noise dominates, and thus using the background error alone dramatically underestimates the true error in the reported flux density. Here we compute and report the full photometric errors from SExtractor for the SpIES survey, correcting for the \Spitzer image flux units such that both background and Poisson noise are included in the error estimate. Indeed the majority of the sources in our ``5$\sigma$ catalog" will have true soure S/N $\textless$ 5 (and more typically $\sim$2-3).

To properly incorporate \Spitzer data into \mbox{Equation \ref{SEerr}}, we first examine its fundamental components: the noise due to the background and Poisson counting noise. In order to compute the background noise, SExtractor first creates a background map and a background rms map. The background rms map is constructed by calculating the squared rms deviation of each pixel in the background map from the local mean background (whose size is defined by the BACK\_SIZE parameter in Table \ref{Params}). The background noise is simply the sum of the background rms pixels inside a given aperture (where $A\sigma_{rms}^2$ in Equation \ref{SEerr} is synonymous with the sum over the background rms). 

Poisson noise is the discrete counting error which occurs when performing photometry on a source. SExtractor performs photometry on an object inside of an aperture by counting the total pixel value and subtracting the background as follows:
\begin{equation}\label{eq:flux}
F=C-B
\end{equation}
where $F$ is the background-corrected count value of an object, $B$ is the sum of the local background value in the aperture, and $C$ is the total number of counts in an aperture. Assuming the pixel values in the measurement aperture are uncorrelated (which presents a separate problem that is discussed later in this section), then the error in $F$ can be calculated using the propagation of error equation:
\begin{equation}\label{eq:prop}
\sigma_{F}^2=\left(\frac{\delta F}{\delta C}\right)^2 \sigma_{C}^2 + \left(\frac{\delta F}{\delta B}\right)^2 \sigma_{B}^2
\end{equation}
where $\sigma_{C}$ and $\sigma_{B}$ are the Poisson errors of the total number of counts and background respectively. Taking the derivatives of Equation \ref{eq:flux} and inserting them into Equation \ref{eq:prop}, we obtain:
\begin{equation}
\sigma_F^2=\sigma_{C}^2+\sigma_B^2 .
\end{equation}

The number of electrons measured, the number of counts reported, and the gain are related by:
\begin{equation}\label{eq:gain}
\#e^- = g \times F
\end{equation} 
which has an uncertainty,
\begin{equation}\label{eq:gainerr}
\sigma_{\#e^-}^2= g^2 \times \sigma_{F}^2.
\end{equation}
Poisson statistics dictate that the variance of a discrete value (in this case electron number, $\sigma_{\#e^-}^2$) is equal to that value (the number of electrons counted). We therefore relate the number of electrons to the digital count in Equation \ref{eq:gain} and obtain that the Poisson error for a digital count is:
\begin{equation}\label{eq:Poisson}
\sigma_{F}^2=\frac{\#e^-}{g^2}=\frac{g \times F}{g^2}=\frac{F}{g}.
\end{equation}
This Poisson error (which must have the digital count unit) is the second term in Equation \ref{SEerr}, and is added in quadrature with the rms background error to generate the total source error found in Equation \ref{SEerr}. 

\Spitzer images and SExtractor use two different definitions of the gain. SExtractor is programmed to interpret this parameter as purely the detector gain (which has units of electrons per digital count) whereas \Spitzer images have a definition of gain that includes the conversion factor between counts units and physical units. Even though SExtractor expects an image in counts units, we can input \Spitzer images by incorporating this conversion factor in the gain parameter according to the equation:
\begin{equation}
G=\frac{N \times g \times T}{K}
\label{gainconv}
\end{equation}
where $N$ is average number of coverages estimated from each AOR coverage map, $g$ is the detector gain of 3.7 $e^-(DN)^{-1}$ for the 3.6\,$\mu$m detector and 3.71 $e^-(DN)^{-1}$ for the 4.5\,$\mu$m detector, $T$ is exposure time for one coverage, and $K$ is the conversion factor from digital to physical units found in either the cBCD header or the Warm IRAC Characteristics webpage\footnote{\href{WarmSpitzer}{http://irsa.ipac.caltech.edu/data/SPITZER/docs/\\irac/warmimgcharacteristics/}}. For the SpIES images, we calculated the weighted gain, $G$, to be 4429.37 $e^-(\mathrm{MJy \,sr}^{-1})^{-1}$ at 3.6\,$\mu$m and 3788.29 $e^-(\mathrm{MJy \,sr}^{-1})^{-1}$ at 4.5\,$\mu$m; these values were used in the SExtractor configuration file for source extraction and error estimation. In short, replacing the detector gain, $g$, with the weighted gain, $G$, in Equation \ref{SEerr} allows a proper determination of both the background and Poisson noise when applying SExtractor to images that have been converted to physical units.

After the gain parameter is replaced, applying simple unit analysis to Equation \ref{SEerr} shows that the errors have the same unit as the input image (in this case MJy sr$^{-1}$). We therefore need to convert the errors from image units of MJy sr$^{-1}$ to $\mu$Jy using Equation \ref{eq:convert} in the same way as we did for the flux density values. The error analysis was also done inside apertures of varying radii and therefore also must be aperture corrected by dividing by the values in Table \ref{tab:aper}. 

Finally, Equation \ref{SEerr} is based on the assumption that the pixels in the images are uncorrelated, which simplifies the SExtractor error calculation. In reality, the SpIES images will have cross correlation terms due to processes such as dithering, reprojection, and stacking, which correlate the count value in overlapping pixels. Since SExtractor does not take correlated noise into account, we corrected the values by multiplying the errors by a factor of two (the ratio of the pre-processed image pixel scale of 1$\farcs$2 to the post-processed pixel scale of 0$\farcs$6), which accounts for the pixels being sampled twice due to the two dithers in the survey. Although the errors are slightly adjusted to account for oversampling, they should still be considered as lower limits on the true error in each aperture since there are other contributions to the correlated noise in each pixel for which we do not correct (i.e.,\ noise pixels). These photometric error estimates will be used in Section \ref{sec:depth} as one of the ways we measure the depth of the survey.


\begin{deluxetable}{ll}
\tablecolumns{2}
\tablewidth{0pt}
\tabletypesize{\scriptsize}
\tablecaption{SpIES catalog columns \label{tab:cols}}
\tablehead{
\colhead{Column Name} & \colhead{Description}}
\startdata
RA\_ch1				&	J2000 RA position at 3.6\,$\mu$m \\
DEC\_ch1				&	J2000 DEC position at 3.6\,$\mu$m \\
FLUX\_APER\_1\_ch1		&	3.6\,$\mu$m flux density, 1$\farcs$44 radius	\\			
FLUX\_APER\_2\_ch1		&	3.6\,$\mu$m flux density, 1$\farcs$92 radius 	\\	
FLUX\_APER\_3\_ch1		&	3.6\,$\mu$m flux density, 2$\farcs$89 radius 	\\
FLUX\_APER\_4\_ch1		&	3.6\,$\mu$m flux density, 4$\farcs$08 radius 	\\
FLUX\_APER\_5\_ch1		&	3.6\,$\mu$m flux density, 5$\farcs$76 radius 	 \\	
FLUX\_APER\_6\_ch1		&	3.6\,$\mu$m flux density, 12$\arcsec$ radius 	 \\		
FLUXERR\_APER\_1\_ch1		&	3.6\,$\mu$m flux density error, 1$\farcs$44 radius   \\
FLUXERR\_APER\_2\_ch1		&	3.6\,$\mu$m flux density error, 1$\farcs$92 radius \\		
FLUXERR\_APER\_3\_ch1		&	3.6\,$\mu$m flux density error, 2$\farcs$89 radius   \\		
FLUXERR\_APER\_4\_ch1		&	3.6\,$\mu$m flux density error, 4$\farcs$08 radius  \\		
FLUXERR\_APER\_5\_ch1		&	3.6\,$\mu$m flux density error, 5$\farcs$76 radius   \\		
FLUXERR\_APER\_6\_ch1		&	3.6\,$\mu$m flux density error, 12$\arcsec$ radius   \\		
FLUX\_AUTO\_ch1			&	Total 3.6\,$\mu$m flux density  \\
FLUXERR\_AUTO\_ch1		&	Total 3.6\,$\mu$m flux density error   \\
FLAGS\_ch1				&	3.6\,$\mu$m SExtractor Flags  \\
CLASS\_STAR\_ch1			&	3.6\,$\mu$m morphology classification \\
FLAG\_2MASS\_ch1			&	3.6\,$\mu$m object near a bright star \\
COV\_ch1 &   Number of cBCD coverages \\
HIGH\_REL\_ch1 & Most reliable objects with good flags \\ 
RA\_ch2				&	J2000 RA position at 4.5\,$\mu$m\\
DEC\_ch2				&	J2000 DEC position at 4.5\,$\mu$m\\
FLUX\_APER\_1\_ch2		&	4.5\,$\mu$m flux density, 1$\farcs$44 radius  \\												
FLUX\_APER\_2\_ch2		&	4.5\,$\mu$m flux density, 1$\farcs$92 radius  \\												
FLUX\_APER\_3\_ch2		&	4.5\,$\mu$m flux density, 2$\farcs$89 radius  \\												
FLUX\_APER\_4\_ch2		&	4.5\,$\mu$m flux density, 4$\farcs$08 radius  \\												
FLUX\_APER\_5\_ch2		&	4.5\,$\mu$m flux density, 5$\farcs$76 radius  \\	
FLUX\_APER\_6\_ch2		&	4.5\,$\mu$m flux density, 12$\arcsec$ radius  \\
FLUXERR\_APER\_1\_ch2		&	4.5\,$\mu$m flux density error, 1$\farcs$44 radius   \\		
FLUXERR\_APER\_2\_ch2		&	4.5\,$\mu$m flux density error, 1$\farcs$92 radius 	 \\		
FLUXERR\_APER\_3\_ch2		&	4.5\,$\mu$m flux density error, 2$\farcs$89 radius 	 \\		
FLUXERR\_APER\_4\_ch2		&	4.5\,$\mu$m flux density error, 4$\farcs$08 radius   \\		
FLUXERR\_APER\_5\_ch2		&	4.5\,$\mu$m flux density error, 5$\farcs$76 radius  \\	
FLUXERR\_APER\_6\_ch2		&	4.5\,$\mu$m flux density error, 12$\arcsec$ radius  \\	
FLUX\_AUTO\_ch2			&	Total 4.5\,$\mu$m flux density   \\
FLUXERR\_AUTO\_ch2		&	Total 4.5\,$\mu$m flux density error  \\		
FLAGS\_ch2				&	4.5\,$\mu$m SExtractor Flags \\    	    
CLASS\_STAR\_ch2			&	4.5\,$\mu$m morphology classification \\	    	     	  	     	       	       			     	  	     	       	       		
FLAG\_2MASS\_ch2			&	4.5\,$\mu$m object near a bright star \\
COV\_ch2 &   Number of cBCD coverages at 3.6\,$\mu$m \\
HIGH\_REL\_ch2 & Most reliable objects with good flags
\enddata
\tablecomments{Column descriptions for the three SpIES catalogs. The 3.6\,$\mu$m-only and 4.5\,$\mu$m-only catalogs are built in exactly the same manner without the columns from the other channel. All flux density and flux density error columns in this catalog have been converted from MJy sr$^{-1}$ to $\mu$Jy pixel$^{-1}$ using Equation \ref{eq:convert}, and the first five apertures in each channel have been aperture corrected using the values in Table \ref{tab:aper}.}
\end{deluxetable}

\subsection{SpIES Source Catalogs}\label{fullcat}

Using the parameters in Table \ref{Params} and employing the techniques discussed in previous sections, we generated the SpIES 5$\sigma$ detection catalogs. Here 5$\sigma$ refers not to objects with a ratio of flux density to flux density error of greater than five, but rather to objects whose flux density is greater than five times the background. This limit is found by taking the product of the DETECT\_MINAREA (minimum number of adjacent pixels to make a source) and DETECT\_THRESH (number of standard deviations above the background per pixel) parameters (see \mbox{Table \ref{Params}} for reference). In fact, the majority of these objects have a S/N of $\sim$2-3, due in large part to the addition of the Poisson noise as shown in Section \ref{sec:errors}.

With this release, we provide three separate detection catalogs: a 3.6\,$\mu$m-only detection catalog which contains $\sim$6.1 million objects that are only detected at 3.6\,$\mu$m, a 4.5\,$\mu$m-only detection catalog containing $\sim$6.6 million objects only detected at 4.5\,$\mu$m, and a dual-detection catalog containing $\sim$5.4 million sources, comprised of the sources detected at the same positions in both bands. These catalogs were constructed by extracting sources from the 3.6\,$\mu$m and 4.5\,$\mu$m AORs separately to generate full object catalogs for each channel. We then matched these two single-band catalogs using a matching radius of $1\farcs3$ (as determined by the Rayleigh criterion), which maximized the number of true matches and minimized the false detections ($\sim$6.5\% for the high reliability objects described below) between the two channels to create our combined dual-band catalog. The objects that did not match remained in the single band catalogs. Due to the offset between the detectors in IRAC, there were $\sim$600,000 objects in 3.6\,$\mu$m without coverage in 4.5\,$\mu$m and $\sim$600,000 objects in 4.5\,$\mu$m without coverage in 3.6\,$\mu$m. These objects, however, are retained in their respective single band catalogs. As the majority of the objects in the single-band catalogs have S/N$\sim$2-3, it is perhaps not surprising that they are detected in only one band.  However, included among these will be transient objects and mid-infrared/optical dropouts, which are clearly of interest, in addition to spurious sources, which are not.  Thus, we recommend using the high reliability flags for the most reliable objects in each catalog (described below).

These catalogs were constructed from the combined epoch AORs, and thus reach the full depth achievable by the SpIES survey. As also noted in the previous section, each row in the catalogs contains a unique source. The columns hold information about the astrometric and photometric values for each source, the flags that were generated during source extraction, and several binary columns which have various meanings (see Table \ref{tab:cols}). The three catalogs are structured in exactly the same way, the only difference being whether or not the object in the catalog is matched between the two channels. A user desiring \emph{all} the 3.6\,$\mu$m detections can concatenate the 3.6\,$\mu$m-only and the dual-band catalogs without any changes to the files.

\begin{deluxetable}{rl}[h!]
\tablecolumns{2}
\tablewidth{0pt}
\tablecaption{Sextractor flags}
\tablehead{
\colhead{Bit} & \colhead{Description}\\ \colhead{Value} & \colhead{}}
\startdata
1   &  The object has neighbors, that significantly bias \\
      & the photometry, or bad pixels. \\
2   &  The object was originally blended. \\
4   &  At least one pixel is (nearly) saturated. \\
8   &  The object is truncated (close to image boundary). \\
16  &  Aperture data are incomplete or corrupted. \\
32  &  Isophotal data are incomplete or corrupted. \\
64  &  A memory overflow occurred during deblending. \\
128 &  A memory overflow occurred during extraction. 
\enddata
\tablecomments{All of the extraction flags from SExtractor. The first five flags are the most common for SpIES as these pertain to issues in source extraction. The last three do not appear in the SpIES data since there are no isophotal aperture measurements and a sufficient amount of memory was allocated for extraction.}
\label{flags}
\end{deluxetable}

Each row in the catalog contains information about a unique source at a particular J2000 RA and DEC position, which was determined by SExtractor, as reported in the first two columns (both channel positions are reported for matched objects). These positions have been corrected for a slight offset when compared to SDSS point sources (see Section \ref{sec:reliability} for more details). The subsequent twelve columns report the flux density values from the six different measurement apertures used in source extraction along with their respective errors. Aperture-corrected flux density values are reported in these columns (except for aperture 6 which is not corrected) and surface brightness units (MJy sr$^{-1}$) are converted to flux densities ($\mu$Jy) using Equation \ref{eq:convert}. Additionally, the errors have been adjusted in the manner described in the previous section. The next two columns (FLUX\_AUTO and FLUXERR\_AUTO) report the flux density and flux density error in apertures whose size and shape are determined by SExtractor to contain the total flux density from a source. These last two values have been converted to flux densities using Equation \ref{eq:convert}; however, they are not aperture corrected. 

\begin{figure}[t!]
\centering
  \includegraphics[width=8cm, height=8cm, trim=5mm 5mm 5mm 5mm,]{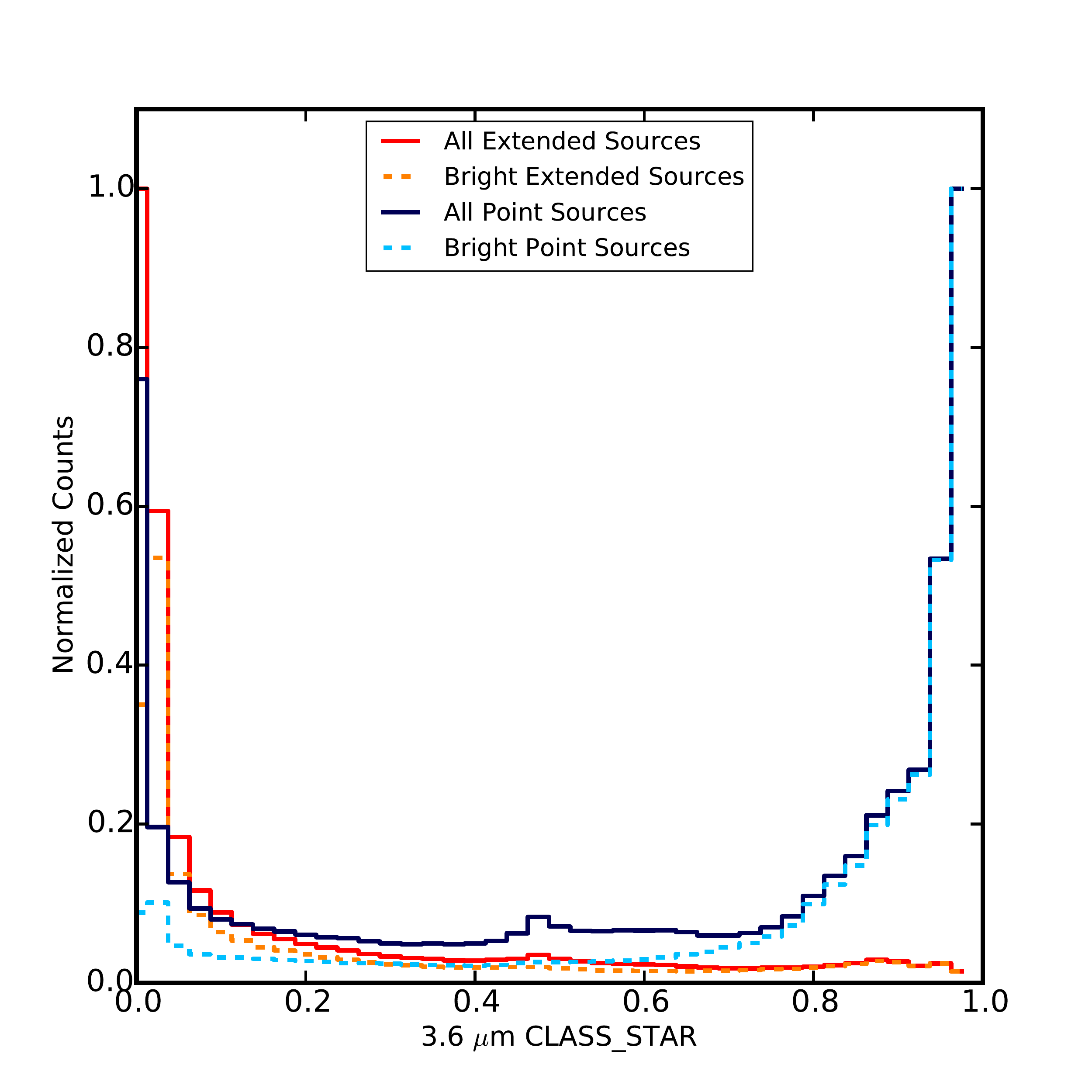}
  \caption{Comparisons of the CLASS\_STAR parameter at 3.6\,$\mu$m for objects matched to SDSS sources. We show the distribution for all optically extended sources (red) and all optical point sources (dark blue). Optically extended sources peak at CLASS\_STAR$\sim$0, while optical point sources peak at $\sim$1; however there is a small peak at 0.5 implying that SExtractor could not differentiate between point or extended. For bright objects ([3.6] $\le$ 20.5), however, the extended (orange dashed) and point (light blue dashed) sources still peak at 0 and 1, respectively, but there are far fewer confused classifications.  A similar trend occurs for the objects detected at 4.5\,$\mu$m.}
\label{fig:class_star}
\end{figure}

The extraction flags are reported in the next column as a 2-dimensional array (see Table \ref{flags} for more information). Since source extraction was performed on an individual AOR basis, the sources on the edges of AORs have the potential to be detected twice, due to the overlap between AORs, and thus both flags were retained (however there is only one row entry in the catalog for overlapping objects). Sources that do not overlap have a flag value in the first array element and were given a value of $-999.0$ in the second element in this column to make it clear that this source was detected in only one AOR. 

The SExtractor stellar class is reported in the CLASS\_STAR column which is a probability that ranges from 0 to 1 and indicates whether an object is resolved (values closer to 1) or extended (values near 0). If the object was detected twice due to the overlap of the AORs, the average value is given in the catalog. We find this measurement to be most reliable for objects with magnitudes brighter than 20.5 ($\sim$1.7 million at 3.6\,$\mu$m and $\sim$1.5 million at 4.5\,$\mu$m in the dual-band catalog), with $\sim$40\% classified as resolved (CLASS\_STAR $\geq$ 0.5) and $\sim$60\% as extended (CLASS\_STAR $\leq$ 0.5) in both bands (see Figure \ref{fig:class_star}).

Following the SExtractor output columns are a series of flags created after source extraction. The FLAG\_2MASS column indicates whether a source is detected within a particular radius (defined by Table \ref{tab:bstar}) around a bright star in the 2MASS point source catalog (PSC). Inside this radius there is an excess of artificial sources due to artifacts from the bright star (e.g.,\ diffraction spikes). Flags are assigned to objects near 2MASS stars with $K_\mathrm{s}$-magnitude brighter than 12 (Vega magnitude), where the radii range from 40$\arcsec$ at the faint end to 180$\arcsec$ at the bright end. For comparison, the radii used for the SWIRE survey range from 10$\arcsec$ at the faint end to 120$\arcsec$ at the bright end using similar (but not the same) $K_\mathrm{s}$-magnitude cuts (see \citealt{Surace2005}). 

\begin{figure}[t!]
\centering
  \includegraphics[width=9cm, height=9cm, trim=5mm 5mm 5mm 5mm,]{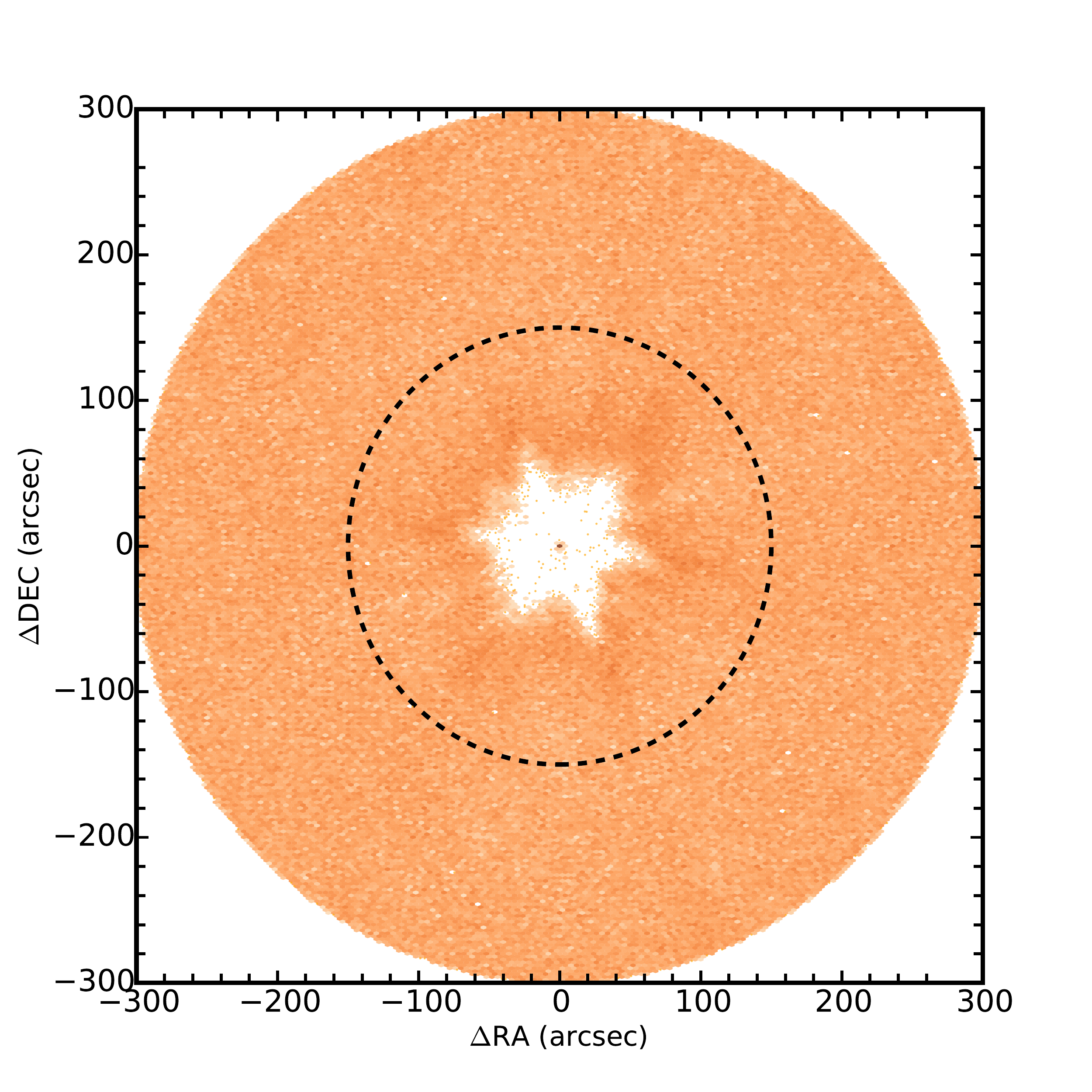}
  \caption{The 335 stacked $6 \le$ $K_\mathrm{s}$-magnitude$\le 7$ stars matched to SpIES within 300$\arcsec$. The black dashed circle shows the radius out to which we flag objects as potentially contaminated.}
\label{fig:maskrad}
\end{figure}

\begin{figure}[h!]
\centering
  \includegraphics[width=9cm, height=9cm, trim=5mm 5mm 5mm 5mm,]{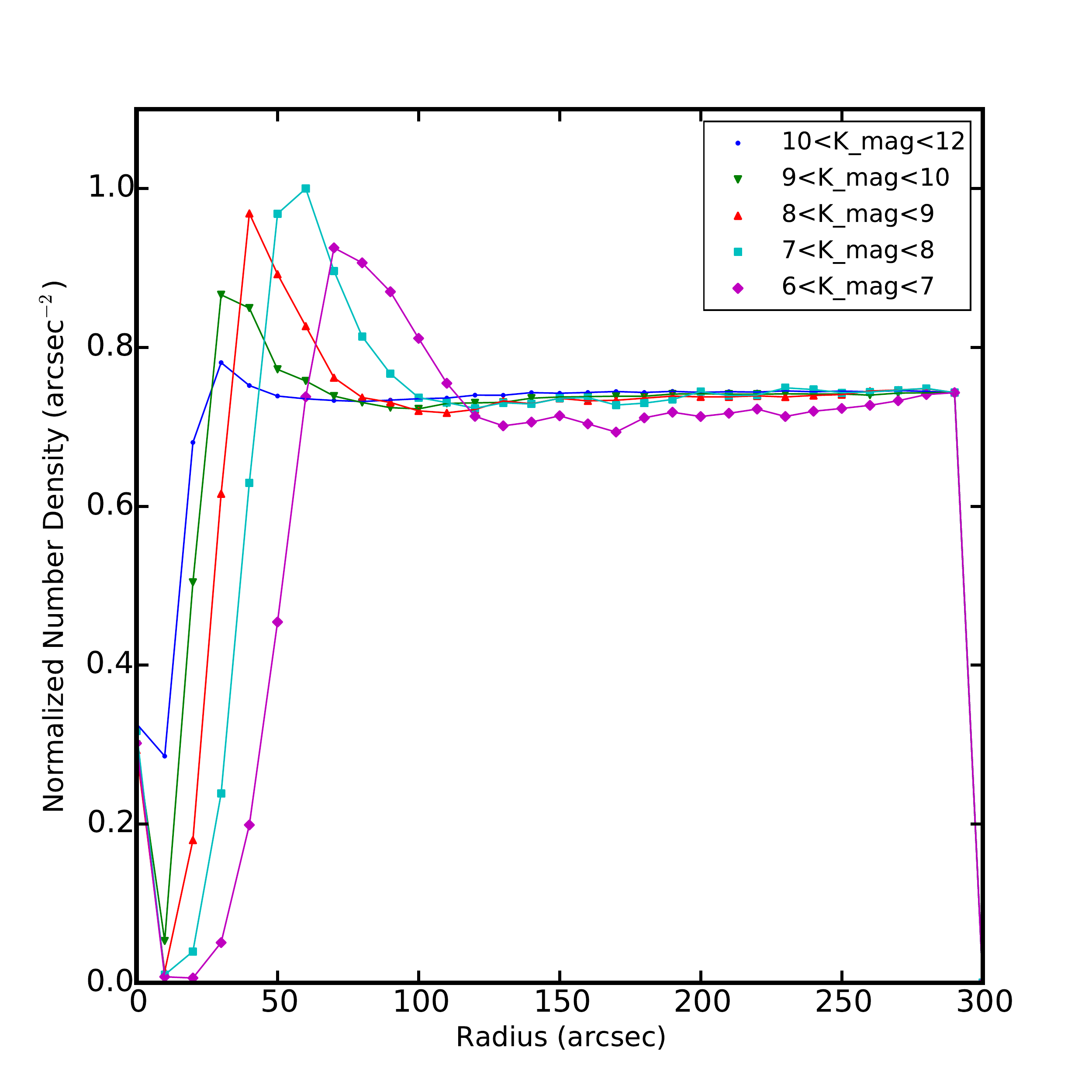}
  \caption{Radial profiles of the number density of objects within 300$\arcsec$ of 2MASS stars in magnitude ranges given in Table \ref{tab:bstar}, showing how the number density of detected objects around bright stars changes as a function of distance from the center of the star. The peak in these curves is the over-dense region where there are spurious detections due to artifacts such as diffraction spikes. We cut at the radius where the curves approach a constant value of number density for each magnitude.}
\label{fig:maskradprof}
\end{figure}

The SpIES bright-star flagging radii were empirically determined by cutting the 2MASS PSC into a series of $K_\mathrm{s}$-band magnitude ranges and matching their positions to all SpIES objects within 300$\arcsec$. We then overlay the positions of all of the stars in a $K_\mathrm{s}$-magnitude bin along with their SpIES matches onto a common coordinate frame and determine the radius which encapsulates the over-dense region around the star. Figure \ref{fig:maskrad} shows the result of stacking $6 \le K_\mathrm{s} \le 7$ Vega magnitude stars and their matches on a coordinate frame. The radial profile plot is presented in Figure \ref{fig:maskradprof} which clearly shows an excess of detections near bright stars. Objects that fall within the radii in Table \ref{tab:bstar} are given a value of $1$ in the catalog to indicate that the source is potentially spurious, and the central star itself is given a value of $2$. Using the radii in Table \ref{tab:bstar}, we compute the area lost when rejecting such sources is $\sim$5 deg$^{2}$ for both bands (which is $\sim$5\% of the dual-band catalog area). 

\begin{deluxetable}{cr}
\tablecolumns{2}
\tablewidth{0pt}
\tablecaption{Bright star flagging radius}
\tablehead{
\colhead{2MASS} & \colhead{Radius}\\
\colhead{($K_\mathrm{s}$-Magnitude)} & \colhead{($\arcsec$)}}
\startdata
$\geq12$ & 0	\\
$12 - 10$ & 40	\\	
$10-9.0 $& 60	\\
$9.0-8.0$ & 90	\\	
$8.0-7.0$ & 120	\\
$7.0-6.0$ & 150	\\
$\leq 6.0$ & 180	
\enddata
\tablecomments{Objects that fall within the radii given are flagged as bright star contaminants. These values are empirically determined by making $K_\mathrm{s}$-magnitude cuts on 2MASS stars and studying figures like Figure \ref{fig:maskrad} and Figure \ref{fig:maskradprof}. The $K_\mathrm{s}$-magnitudes are computed in Vega magnitudes.}
\label{tab:bstar}
\end{deluxetable}

We report the number of cBCD coverages (from the coverage maps shown in Figure \ref{AORpic}) at the centroid position of each source in the COV column. Since AORs overlap, we give an array of two values where, if the object does not overlap, we report $-$999.0 in the second element (similar to the extraction flags). For the most reliable detection, we recommend using objects which have greater than two coverages in either entry of the reported array. 

Finally, we have created a high reliability column which we recommend for users whose science requires that the objects be robust sources and/or have robust photometry. There are three values in this column indicating whether a source is a real object (flagged with a value of 1 or 2), has good photometry (flagged with a value of $2$), or does not satisfy the following good flag conditions (flagged with a $0$). To be regarded as a real source, the SExtractor flags must be less than or equal to $4$, the objects must have flag $0$ or $2$ in the FLAG\_2MASS column, and there has to be greater than or equal to two coverages for each source. For an object to have good photometry, we further require that the SExtractor flags be less than or equal to 0 or equal to 2 (i.e.,\ $-$999.0, 0, and 2), FLAG\_2MASS must be $0$, and it must satisfy the same coverage conditions as before. These flags cause holes in the coverage across the survey, thus changing the total coverage area. In total, SpIES has $\sim$115 deg$^{2}$ of coverage in both wavelengths, of which, each band covers $\sim$107 deg$^{2}$ (since there is an offset in the arrays discussed in Section \ref{sec:Obs_Strat}) and there is $\sim$100 deg$^{2}$ of dual-band coverage. For HIGH\_REL$\textgreater$0, the areas are $\sim$106 deg$^{2}$, $\sim$101 deg$^{2}$, and $\sim$94 deg$^{2}$, while for HIGH\_REL$=$2, the areas drop to 105 deg$^{2}$, $\sim$100 deg$^{2}$, and $\sim$89 deg$^{2}$. While our catalog only includes sources more than 5$\sigma$ above the background, full error analysis means that individual objects can have S/N (as computed by FLUX/FLUXERR) less than 5. Some users may want to apply a cut on S/N in addition to using the HIGH\_REL flag. For a cut at S/N$\textgreater$3 and HIGH\_REL$\textgreater$0, we retain $\sim$1.4, $\sim$3.9, and $\sim$1.4 million objects in the 3.6\,$\mu$m-only, dual-band, and 4.5$\mu$m-only, respectively. 

\subsection{Astrometric Reliability}\label{sec:reliability}

\begin{figure}
\begin{center}
  \includegraphics[width=8cm, height=8cm, trim=5mm 5mm 5mm 5mm]{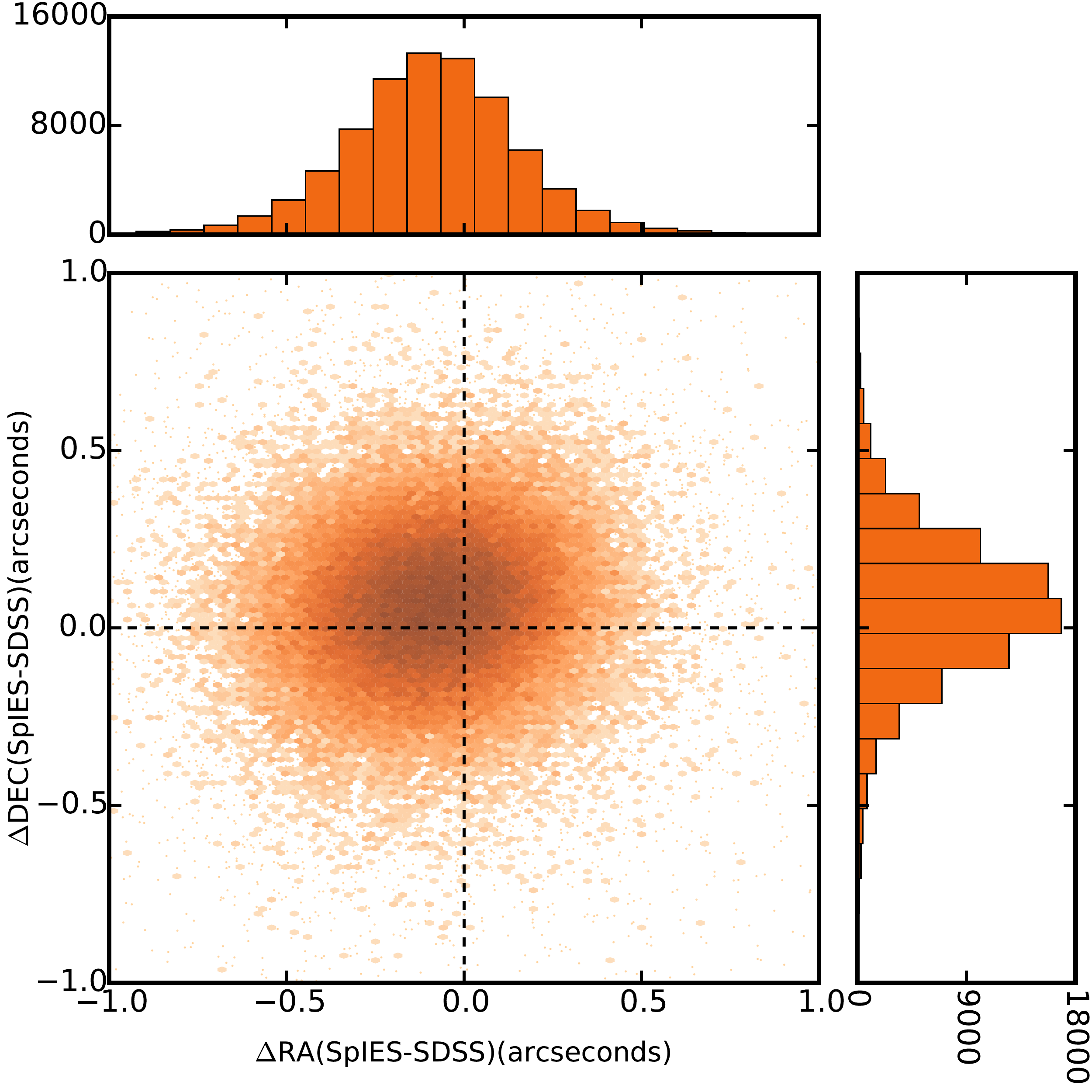}
\caption{Comparison of the SpIES and SDSS astrometry for matched point sources with good flags in both surveys. Darker regions and histograms show the approximate point density. We use the mean offsets of the $\Delta$RA and $\Delta$DEC distributions shown here to correct the SpIES astrometry.}
\label{fig:astrometry}
\end{center}
\end{figure}

The astrometric reliability of SpIES was tested by comparing the centroid positions of point sources in SDSS with matched objects in the SpIES dual-band catalog (within 2$\arcsec$). We found the difference in position for objects which have good flags in SDSS (BITMASK=0 and PHOTOMETRIC=1), are bright in the $r$-band ($r \le$ 21), and have good flags in SpIES (HIGH\_REL=2). Fitting a Gaussian to the histograms in Figure \ref{fig:astrometry}, we find that the mean difference in RA is $-0\farcs112 \pm 0\farcs0008$ and in DEC is $0\farcs0372 \pm 0\farcs0006$ for these objects. These values were then used to correct the astrometry in all three SpIES catalogs. We also matched the SpIES data with the 2MASS PSC and found that the mean astrometric offsets ($\Delta$RA$=-0\farcs086 \pm 0\farcs0006$ and $\Delta$DEC$=0\farcs011 \pm 0\farcs0005$) are slightly smaller than the calculations from SDSS, however confirm the direction of the SpIES positional shifts. 

To see if the astrometric offset changes with brightness, we performed the same measurement using the SDSS matched point sources for bright and faint sources in [4.5]. We find that the astrometric offsets to be rather consistent both for faint ([4.5] $\ge$ 20 mag) objects with $\Delta$RA$=-0\farcs112 \pm 0\farcs0009$ and $\Delta$DEC$=-0\farcs0370 \pm 0\farcs0007$ and for bright objects ([4.5] $\le$ 20 mag) with $\Delta$RA$=-0\farcs112 \pm 0\farcs0014$ and $\Delta$DEC$=-0\farcs0376 \pm 0\farcs0012$. Regardless of magnitude, with the 0$\farcs$6 pixel scale of the SpIES images, the astrometric offset is approximately one sixth of a pixel, which is similar to the values calculated in \citet{Ashby2009} where the SDWFS astrometry was compared to 2MASS. 

\begin{figure*}
  \includegraphics[width=9cm, height=9cm, trim=5mm 5mm 5mm 5mm, clip]{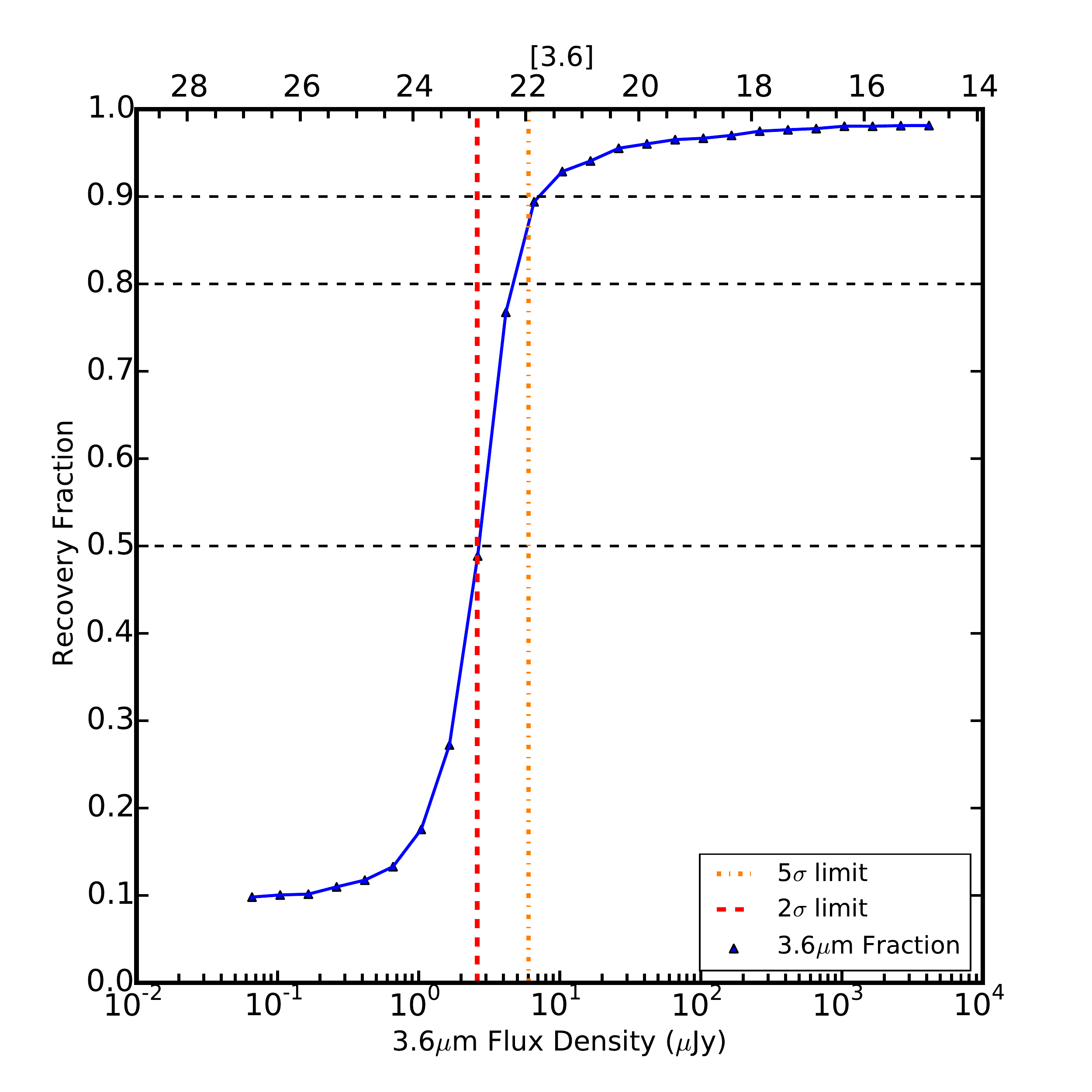}
  \includegraphics[width=9cm, height=9cm, trim=5mm 5mm 5mm 5mm, clip]{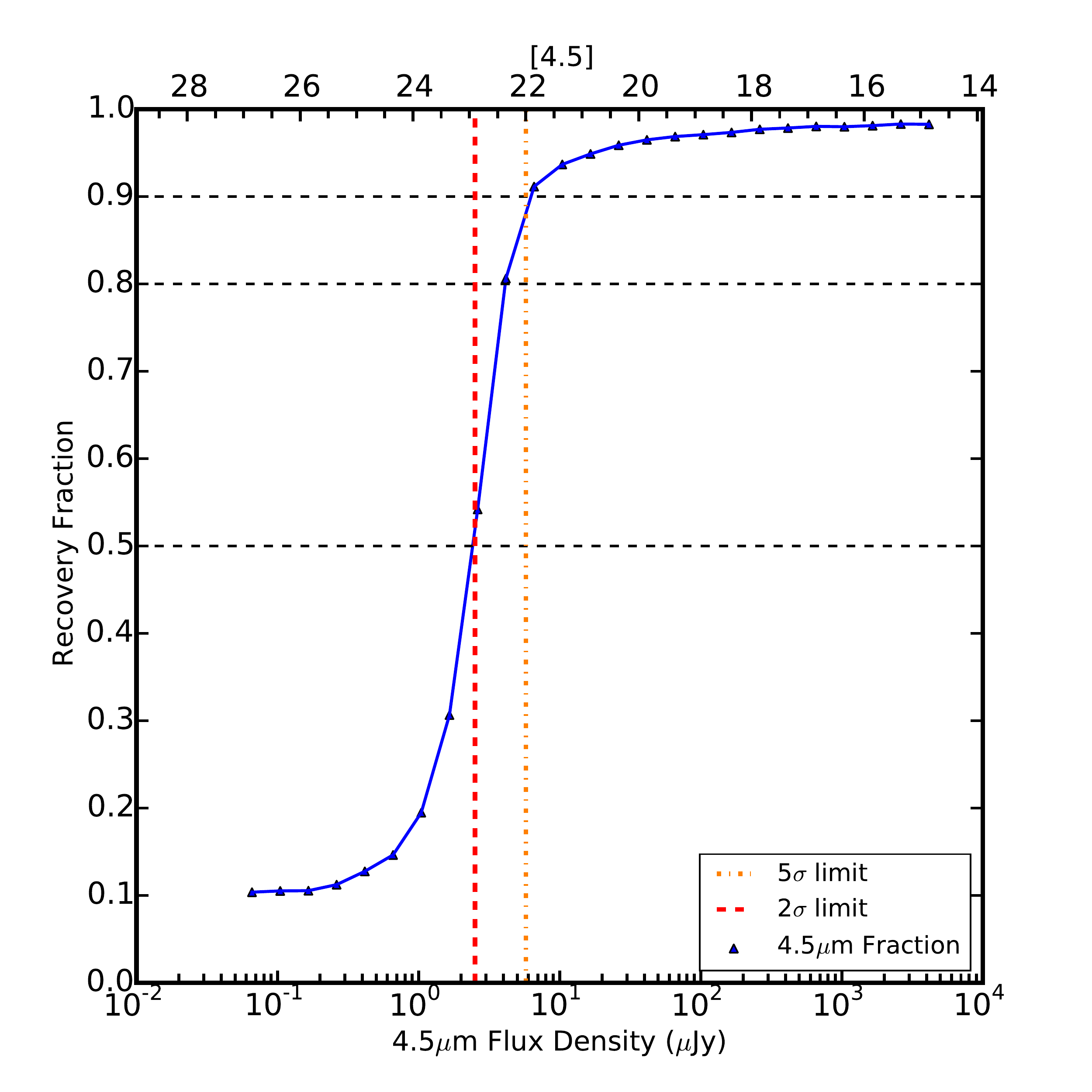}
\caption{Completeness as a function of 3.6\,$\mu$m flux density (and [3.6]; left) and 4.5\,$\mu$m flux density (and [4.5]; right) of our simulated sources. The orange dot-dashed line marks the faintest detection of (5$\sigma$) objects at 6.13 $\mu$Jy and 5.75 $\mu$Jy at 3.6\,$\mu$m and 4.5\,$\mu$m, respectively; the red dashed line shows (2$\sigma$) objects at 2.58$\mu$Jy and 2.47$\mu$Jy at 3.6\,$\mu$m and 4.5\,$\mu$m, respectively, as measured from the curves in Figure \ref{Depthfig}. The completeness curves are less affected by artifacts at faint magnitudes since the analysis is done with simulated sources, and thus are better estimates of depth than the number counts.}
\label{fig:Completeness}
\end{figure*}  

 \subsection{Completeness and Number Counts}\label{Complete}
To estimate the completeness of our detection strategy, we employed a Monte Carlo approach where we simulated 15,000 sources (between 4$\%$ and 6$\%$ of the total number of sources) with random magnitudes between 14.5 and 28 at random positions on each AOR. The simulated sources were allowed to fall anywhere on the image, including on top of other sources, thus our completeness estimates are robust against confusion noise (see \citealt{Ashby2013}). Each source was modeled as a point source, having a Gaussian profile with the same FWHM as IRAC. We ran SExtractor on these simulations in the exact manner described in Section \ref{Extraction} and matched to a file containing the position and magnitude for each source. The tables of recovered sources for each AOR were then concatenated as before to cover the full footprint of SpIES. Number counts as a function of magnitude were plotted for both the recovered object catalog and the full simulated source catalog and the ratio of counts in each bin was calculated to estimate the completeness of the survey. Figure \ref{fig:Completeness} presents the SpIES completeness curve for each passband, and the 90, 80, and 50 percent completeness values are quoted in Table \ref{Complevel}. These measurements are performed for the entire survey field, however SpIES spans a wide range in right ascension. We therefore evaluated the completeness at different ranges in right ascension to evaluate how it changes with position. We found that the differences between the completeness curves that were computed for the full survey in Figure \ref{fig:Completeness} and the curves computed at different locations in the SpIES survey were not significantly different, and that the differences in the 90, 80, and 50 percent complete values do not exceed $\sim$0.15 magnitudes for both the 3.6\,$\mu$m and 4.5\,$\mu$m measurements.

\begin{figure*}
  \includegraphics[width=9cm, height=9cm, trim=5mm 5mm 5mm 5mm, clip]{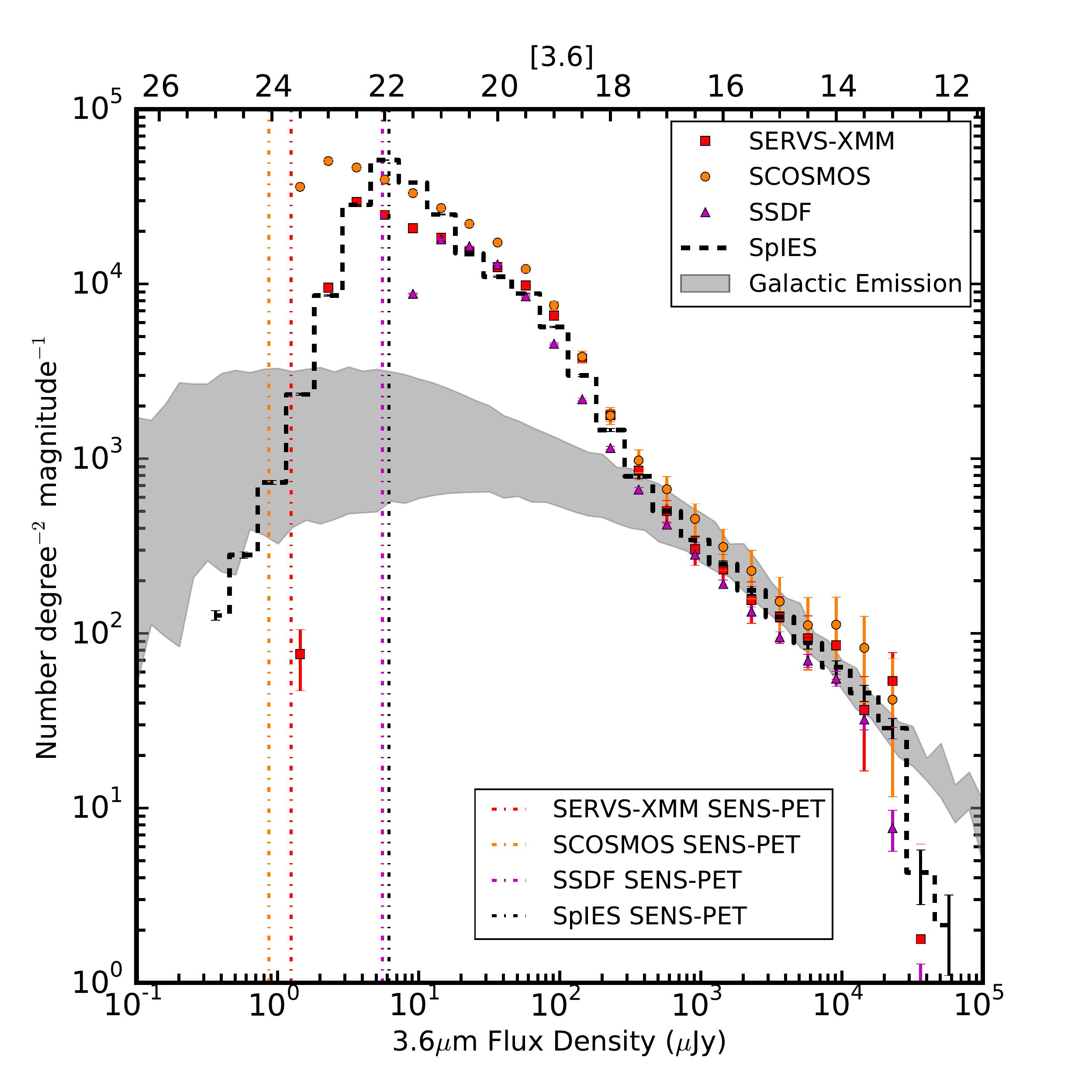}
  \includegraphics[width=9cm, height=9cm, trim=5mm 5mm 5mm 5mm, clip]{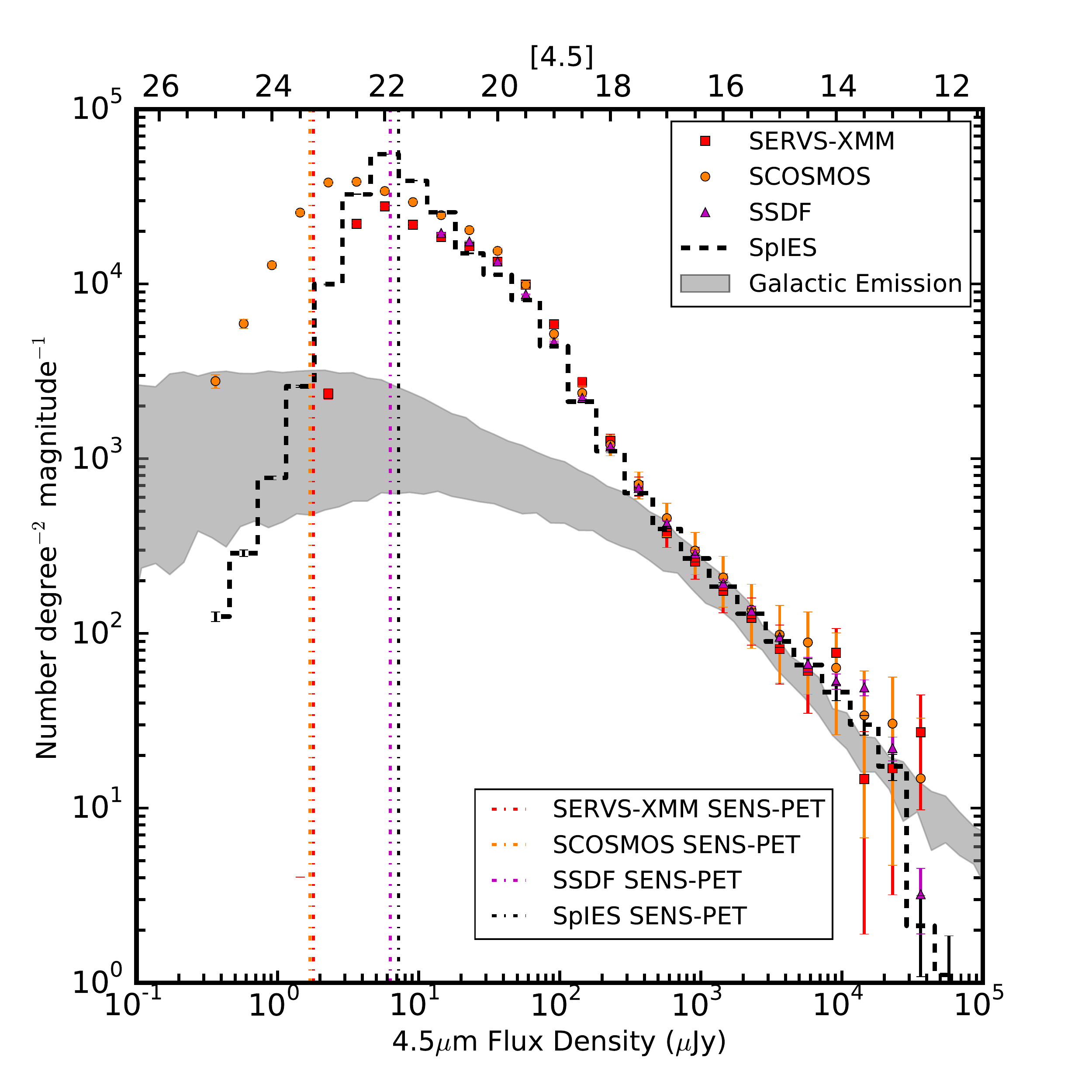}
  \caption{Differential number counts per magnitude over the full SpIES field for all objects with a HIGH\_REL $\textgreater$ 0. In both panels, we divide the counts by an area of 101 deg$^{2}$ which is the area covered for this footprint in each detector. Left: SpIES 5$\sigma$ catalog (black dash) histogram of number of objects per square degree vs flux density ($\mu$Jy) for all objects detected at 3.6\,$\mu$m. Also shown are the number counts from the SERVS XMM field (\citealt{Mauduit2012}; red squares), S-COSMOS (\citealt{Sanders2007}; orange circles), and SSDF (\citealt{Ashby2013}; purple triangles) as comparisons. The vertical dot-dashed lines represent the SENS-PET predicted depth for each survey. As we include objects that are more than 5$\sigma$ above the background, but have S/N $\textless$ 5, the excess relative to other surveys near the 90\% completeness limit is likely an indication of contamination by low probability sources. Right: The 4.5\,$\mu$m number counts similar to the left panel. The grey shaded region shows the contribution of Milky Way stars using the DIRBE Faint Source Model \citep{Arendt1998, Wainscoat1992}.}
\label{fig:Flux_hists_spies_5s}
\end{figure*}

Differential number count histograms provide a visual representation of the distribution of objects of different magnitudes in a survey. They can be used to approximate the number of particular objects (stars, quasars, galaxies, etc.) that should be detected in the survey and can provide a rough estimate of the depth of the survey. The number of objects per square degree per magnitude is plotted as a function of flux density and AB magnitude in Figure \ref{fig:Flux_hists_spies_5s} for SpIES objects detected in each band that satisfy the condition HIGH\_REL$\textgreater$0. Shown for comparison are the differential number counts from SSDF \citep{Ashby2013}, which has a similar depth as SpIES, along with counts from the SERVS XMM field \citep{Mauduit2012} and the S-COSMOS survey \citep{Sanders2007}, both of which are deeper than SpIES. Additionally, we show the contribution of Milky Way stars to these number counts estimated using the DIRBE Faint Source Model (FSM; \citealt{Arendt1998, Wainscoat1992}). At the bright end, the four surveys and the FSM all tend to align and follow a similar linear trend, indicating that the bright objects in the SpIES catalog are well represented and are mostly attributed to light in the Milky Way. The ``turn over" in these histograms indicates the location of the approximate value of the depth of the survey. This is, however, an imperfect measure of the depth since artifacts tend to increase at the faint limits of a survey, resulting in more counts at fainter magnitudes.

The SpIES differential number counts in Figure \ref{fig:Flux_hists_spies_5s} are computed for the full footprint of the survey. The spatial extent of SpIES is large enough, however, that it intersects the Galactic plane at different angles which has a small effect on the number counts, particularly for faint objects (20 $\le$ AB $\le$ 22). For this reason the FSM, which is calculated for only a small area on the sky,  is represented by a grey shaded region. To test the effect of Galactic latitude on the number counts, we split SpIES into different regions at different Galactic latitudes (0$\le$ $b$ $\le$ 15, 15$\le$  $b$  $\le$ 30, and  $b$  $\ge$ 30) and recompute the number counts as a function of magnitude. We find fewer faint objects are recovered for low Galactic latitudes, however as we look further off of the Galactic plane the SpIES number counts become consistent with those for surveys of similar depth (i.e.,\ SSDF). 

\subsection{Depth}\label{sec:depth}
\begin{figure}
\includegraphics[width=9cm, height=9cm, trim=5mm 5mm 5mm 5mm, clip]{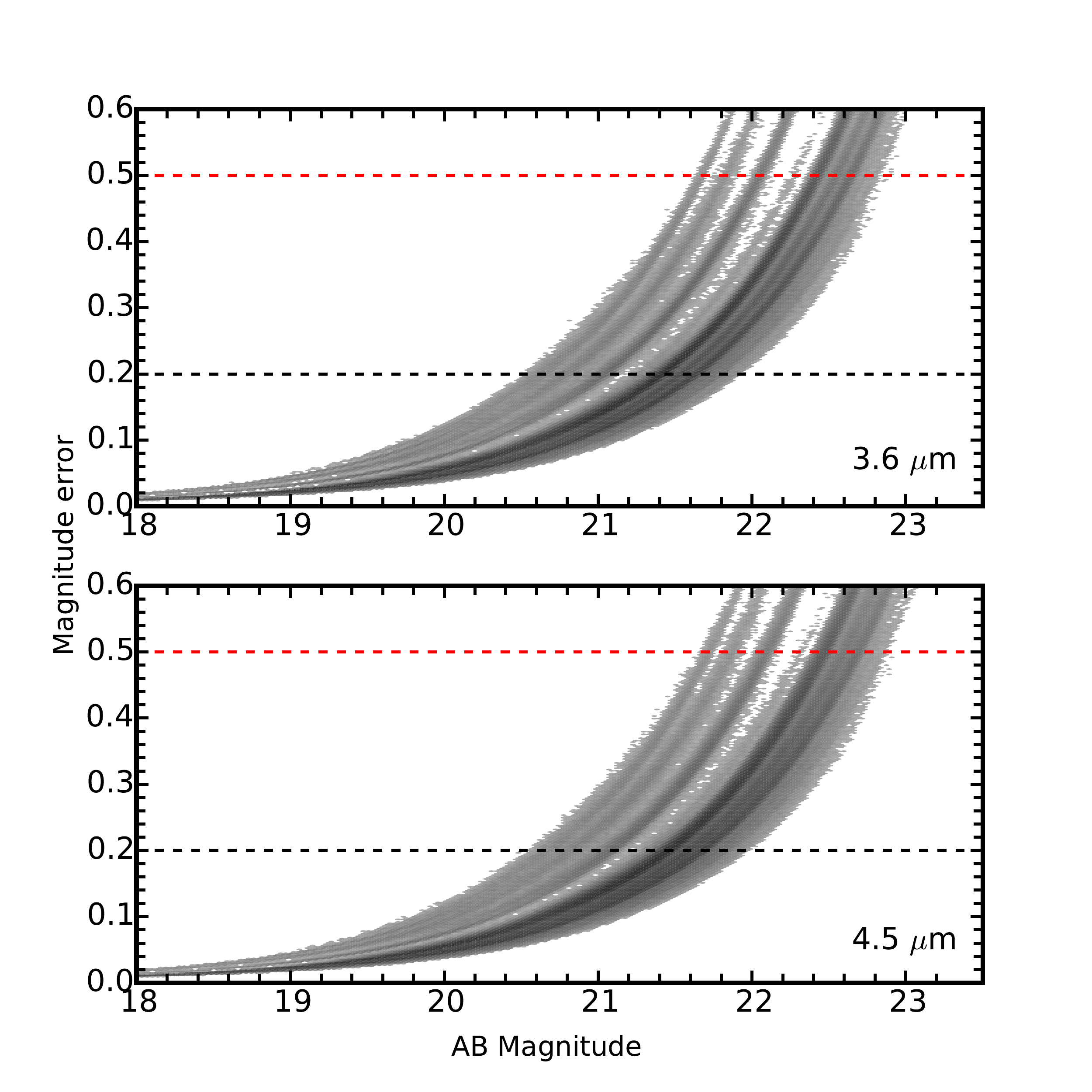}
\caption{Estimation of the SpIES detection limit at 3.6\,$\mu$m (top) and 4.5\,$\mu$m (bottom). The grey points indicate the error in magnitude vs.\ magnitude. The 5$\sigma$ limit occurs at a magnitude error of 0.2 (black dashed line), and the 2$\sigma$ limit occurs at a magnitude error of 0.5 (red dashed line). These values are determined by propagating the error in the expression for magnitude, resulting in the ratio of noise to signal as the error in magnitude. The intersection of the right edge of the grey points with the respective magnitude error is the approximate detection threshold. Differences in shading indicates the density of points.}
\label{Depthfig}
\end{figure}
    
There are multiple ways of determining the depth of a survey, and the optimal value to use depends on the intended application. We computed the depth in four different ways for our analysis. First, we find the magnitude where the completeness curves turn over (see Figure \ref{fig:Completeness}). Object detection declines rapidly at this magnitude, making it a useful indicator of survey depth. An estimate of the limiting magnitude using the 90th percentile of completeness for simulated sources is [3.6]=21.75 and [4.5]=21.90. We report the 90, 80, and 50 percent complete values in Table \ref{Complevel}.

Secondly, we can estimate the 5$\sigma$ and 2$\sigma$ depths by plotting the magnitude error as a function of magnitude (see Figure \ref{Depthfig}). From Figure \ref{Depthfig} we determine the magnitude value where the outer edge of the curve reaches a magnitude error of $\sim$0.2 to obtain the 5$\sigma$ magnitude limit. For SpIES, this limit occurs at [3.6]=21.93 and [4.5]=22.00, which corresponds to flux density values of 6.13 $\mu$Jy and 5.75 $\mu$Jy, respectively. 

Another method to estimate depth is to perform empty aperture photometry where we placed random apertures on the images and performed source extraction in each aperture. We then made a histogram of the measurements with negative flux density values in the 1$\farcs$9 aperture in an attempt to eliminate contamination from sources to the background measurements. We then fit a Gaussian curve to the data to find the standard deviation in the background, $\sigma_{bg}$, across the SpIES field. We find that the 5$\sigma_{bg}$ measurements are 8.14 $\mu$Jy at 3.6\,$\mu$m and 7.55 $\mu$Jy at 4.5\,$\mu$m. While this does not directly measure the depth to which we observe, it is a robust measurement of the noise in the data, including confusion noise since the apertures were randomly placed on our images.

Finally, we use the predicted limits produced by the SENS-PET\footnote{\href{SENSPET}{http://ssc.spitzer.caltech.edu/warmmission/propkit/pet/senspet/}} tool. This estimate calculates the 5$\sigma$ point source depth given the background level of the survey (depending on the survey location), the exposure time, and number of repeat exposures over a single area. The SpIES depth is estimated at 6.15 $\mu$Jy at 3.6\,$\mu$m and 7.2 $\mu$Jy at 4.5\,$\mu$m using a medium background, an exposure time of 30 seconds, and four overlaps in the `Warm IRAC Parameters' section. This tool appears to calculate depths that are shallower than the measured depths; however, it is useful for making robust comparisons to other survey fields (for example, see Figure \ref{fig:Flux_hists_spies_5s}).

There are multiple reasons for the slight differences between the prediction from SENS-PET and our measurements. First, the noise estimates previously discussed in Section \ref{sec:errors} should be considered a lower limit on the error and therefore the signal-to-noise ratios may be overestimated. Second, an overlap value of 4.0 was inserted into the SENS-PET calculator, whereas in reality the overlap of the SpIES BCD images averages to a value of $\sim$4.5 per pixel. The more coverage, the deeper the observations, so the theoretical value will be slightly brighter than reality. Finally, there could be a disparity between the background model used in SENS-PET and the measured background from the SpIES AORs, which could lead to a difference in the depth.

\begin{deluxetable}{ccccc}
\tablecolumns{5}
\tablewidth{0pt}
\tablecaption{Completeness levels}
\tablehead{
\colhead{Level} & \multicolumn{2}{c}{3.6\,$\mu$m}& \multicolumn{2}{c}{4.5\,$\mu$m}}
\startdata
$90\%$ complete&21.75&7.2$\mu$Jy & 21.90&6.3$\mu$Jy	\\
$80\%$ complete&22.20&4.8$\mu$Jy & 22.37&4.1$\mu$Jy	\\
$50\%$ complete& 22.82&2.7$\mu$Jy & 22.91&2.5$\mu$Jy	\\
5$\sigma$ &21.93&6.13$\mu$Jy & 22.00&5.75$\mu$Jy	\\
2$\sigma$ & 22.87&2.58$\mu$Jy & 22.92&2.47$\mu$Jy	\\
5$\sigma_{bg}$ &21.62&8.14$\mu$Jy & 21.70&7.55$\mu$Jy	\\
2$\sigma_{bg}$ & 22.62&3.26$\mu$Jy & 22.70&3.02$\mu$Jy	\\
SENS-PET 5$\sigma$ &21.93&6.15$\mu$Jy &21.76&7.20$\mu$Jy	
\enddata
\tablecomments{We give the 90, 80 and 50 percent completeness levels in AB Magnitudes and flux density of the SpIES survey from Figure \ref{fig:Completeness} as well as the 5$\sigma$ and 2$\sigma$ values from Figure \ref{Depthfig}, the empty aperture measurements at 5$\sigma_{bg}$ and 2$\sigma_{bg}$, and the SENS-PET estimates.}
\label{Complevel}
\end{deluxetable}      

\subsection{Confusion}\label{Conf}
We estimate the threshold for source confusion (the noise attributed to faint or unresolved background sources) by calculating the average number of SpIES beams per source, similar to the technique used in \citet{Ashby2009}, and compare with the classical threshold limits determined in \citet{Condon1974} and \citet{Hogg2001}. The SpIES beam size (solid angle) is calculated using $\Omega=\pi\sigma^2$, where $\sigma$ is the standard deviation of the Gaussian point spread function. Using the relation $\mathrm{FWHM}=2\sqrt{2\mathrm{ln}(2)}\sigma$ and the `warm' IRAC FWHM values of 1$\farcs$95 in the 3.6\,$\mu$m detector and 2$\farcs$02 in the 4.5\,$\mu$m detector, we obtain a beam size of 2.155 arcsec$^2$ for the 3.6\,$\mu$m detector and 2.312 arcsec$^2$ for the 4.5\,$\mu$m detector. The total number of beams over the full SpIES area is $6.92\times10^8$ in the 3.6\,$\mu$m images and $6.45\times10^8$ in the 4.5\,$\mu$m images. Finally, taking the ratio of the number of beams to the number of objects at different detection thresholds yields an estimate for the confusion.

There are a total of $\sim$11.6$\times10^6$ objects detected at 3.6\,$\mu$m (combining the 3.6\,$\mu$m-only catalog and the dual-band catalog) and $\sim$12.1$\times10^6$ objects detected at 4.5\,$\mu$m (combining the 4.5\,$\mu$m-only catalog and the dual-band catalog) before applying flags for known contaminants, thus there are $\sim$60 beams per source and $\sim$53 beams per source for the full 3.6\,$\mu$m and 4.5\,$\mu$m detection catalogs, respectively. Taking the inverse of these two results suggest that approximately $1.6\%$ of the detections at 3.6\,$\mu$m and $1.9\%$ of the detections at 4.5\,$\mu$m are confused. \citet{Condon1974} and \citet{Hogg2001} found the threshold for confusion to be significant when there are fewer than 30 to 50 beams per source for number counts histograms which have power law slopes of 0.75 to 1.5. The SpIES number counts histograms have slopes of $\sim$0.85 for both bands, therefore, with 60 and 53 beams per source at 3.6\,$\mu$m  and 4.5\,$\mu$m, respectively, we conclude that SpIES is not significantly affected by source confusion.

\section{Diagnostics and Summary}\label{sec:sourcematch}
\begin{figure*}[h!]
\begin{center}
  \includegraphics[width=15cm, height=15cm, trim=5mm 5mm 5mm 5mm, clip]{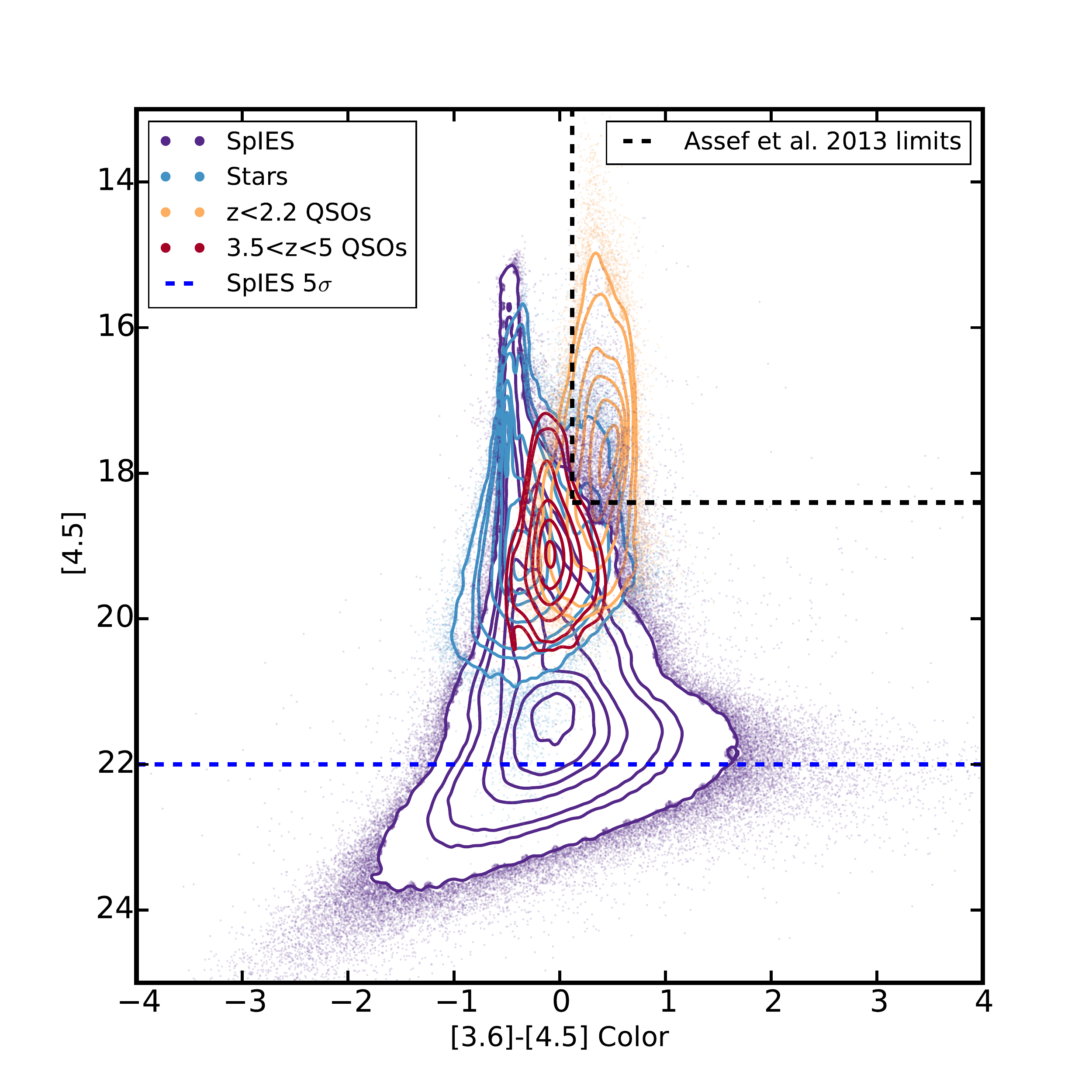}
\caption{Color-magnitude diagram for SpIES objects with good photometry (i.e.,\ HIGH\_REL$=$2; purple). Also indicated are contours of where different objects fall in this color space. The blue contours are stars, light orange contours are known low-redshift quasars $(z\le 2.2)$, and red contours are high-redshift quasars $(3.5 \le z \le 5)$. These additional contours are not objects matched to SpIES data, rather are SDSS detections which have \Spitzer color information. We show the superior depth of the SpIES survey (the blue dashed line is the [4.5]=22.00 5$\sigma$ line) compared to the star and quasar data from the optical. The black dashed lines represent the \citet{Assef2013} criteria for AGN selection in this color space ($W1$-$W2$ $\ge$ 0.8), which, although very complete for low-redshift quasars (obscured and unobscured), misses most high-redshift quasars (e.g.,\ \citealt{Richards2015}). We draw contours which encapsulate 10 to 90 percent of the data (in 20 percent increments) and 95 percent of the data. We additionally draw 99 percent contours for the SpIES objects (purple) and stars (blue).}
\label{fig:colmag_5s}
\end{center}
\end{figure*}

\subsection{Color Distributions}
To test the accuracy of our data processing, we examine the distribution of magnitudes and colors of SpIES sources and compare them to known objects and infrared photometry from \emph{WISE}. Mid-infrared color-color diagrams have proven to be effective in classifying objects, for example quasars, as shown in \citet{Lacy2004}, \citet{Stern2005}, and \citet{Donley2012}. Unlike these previous IRAC analyses, which had access to all four channels, SpIES only observes in the first two, thus instead of a color-color diagram, we investigate the color-magnitude space shown in Figure \ref{fig:colmag_5s}. All SpIES sources with HIGH\_REL$=$2 (in both bands) from the dual-band catalog are shown, along with stars and spectroscopically-confirmed quasars (drawn from the \citet{Richards2015} ``master" quasar catalog) which are detected in both the optical and by \emph{Spitzer}. 

The ``master" catalog is a combination of spectroscopically-confirmed quasars from SDSS-I/II/III \citep{York2000, Eisenstein2011} matched with photometric sources from the AllWISE survey. To the ``master" catalog, we have added new $z \textgreater 5$ quasars from \citet{McGreer2013} and the SDSS DR12 quasar catalog (P\^{a}ris et al.\ 2016, in preparation). The \emph{WISE} Vega magnitudes in the ``master" catalog have been converted to AB magnitudes by adding 2.699 to W$_1$ and 3.339 to W$_2$ which is the difference in the respective zero points for the \emph{WISE} detectors. The \emph{WISE} AB magnitudes were then converted to the \emph{Spitzer} AB system using the method in Section 2.3 of \citet{Richards2015} and Table 1 of \citet{Wright2010}. The \emph{Spitzer} and \emph{WISE} detectors take images at slightly different wavelengths, and therefore observe emission from an object at slightly different locations in its spectral energy distribution. The conversion factor between the two detectors is, therefore, dependent on the color of the observed object. For our analysis, we adopt the look-up table from \citet{Richards2015} which provides the proper correction for an object with a given color and spectral index (assuming a power-law spectral energy distribution). Figure \ref{fig:colmag_5s} demonstrates that SpIES can be used to distinguish various types of objects in the mid-infrared. Stars, for example, appear bluer ([3.6]-[4.5]\textless 0) than low-redshift ($z \le 2.2$) quasars, which tend to lie in a redder ([3.6]-[4.5]\textgreater 0) region of this diagram, despite covering approximately the same magnitude range at 4.5\,$\mu$m. It is also apparent that SpIES is achieving a depth that exceeds that of the spectroscopic quasar sample shown.

\subsection{SDSS quasars}
\begin{figure}
\begin{center}
 \includegraphics[width=9cm, height=9cm, trim=5mm 5mm 5mm 5mm, clip]{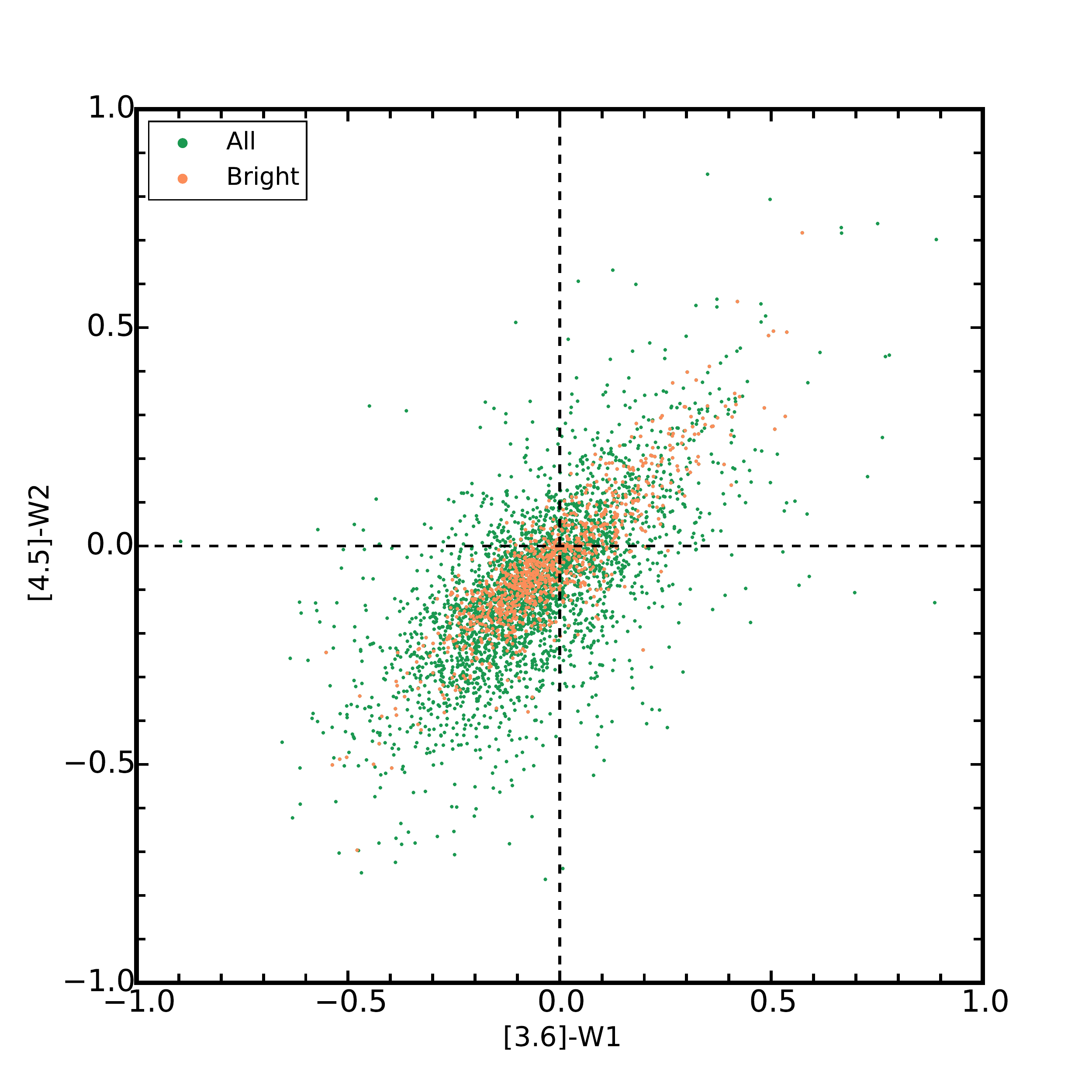}
\caption{Comparison of the  SpIES and WISE colors for quasars from the \citet{Richards2015} ``master" catalog. WISE Magnitudes have been corrected to the IRAC AB Magnitude system in both channels. The orange points show the color of the brightest quarter of the WISE data (W1$\le$15.5 \& W2$\le$15.5 WISE Vega magnitudes). In principle, we expect the points to be near the origin, however phenomena such as variability and systematics such as contamination in \emph{WISE} W1 and W2 cause the points to deviate.}
\label{fig:spieswisecolors}
\end{center}
\end{figure}

Figure \ref{fig:spieswisecolors} displays [3.6]$-$W1 vs [4.5]$-$W2 for the confirmed quasars in the \citet{Richards2015} ``master" quasar catalog. In theory, we might expect the quasar colors to converge at the origin, however there is a deviation of the colors from the origin which can be attributed to a few factors. First, SpIES and the AllWISE surveys were conducted at different times, and thus variable quasars would shift diagonally in this color space. Additionally, there is a well-known flux underestimation bias for fainter objects in the AllWISE data attributed to an overestimation of the background caused by contamination of nearby objects, forcing the \emph{WISE} colors to appear fainter (see the AllWISE Explanatory Supplement\footnote{\href{allwise}{http://wise2.ipac.caltech.edu/docs/release/allsky\\/expsup/sec6\_3c.html\#flux\_under}} for more detail). 
    
One of the goals of SpIES is to uncover new, faint quasars at high-redshift to use for clustering investigations. From Figure \ref{fig:colmag_5s}, it is apparent that cuts in infrared color-magnitude space alone will not cleanly select high-z quasars.  However, quasar candidates can be selected using the multidimensional selection algorithm described in \citet{Richards2015} which analyzed the colors of quasars in the optical with SDSS and infrared with AllWISE. They constructed a training set of quasars comprised of objects in the AllWISE catalog that have spectroscopically confirmed quasar counterparts in SDSS (i.e.,\ known quasars), and a test set comprised of AllWISE objects that have SDSS photometry. Using the colors of the known quasars in the training set as a Bayesian prior, probabilities were assigned to the objects in the test set based off of where they lie in the optical-infrared, multidimensional color space. We will follow this technique using, the SpIES data instead of AllWISE since it probes much deeper and has superior resolution, allowing us to better select high-redshift quasar candidates on S82.

\begin{figure}
\begin{center}
 \includegraphics[width=9cm, height=9cm, trim=5mm 5mm 5mm 5mm, clip]{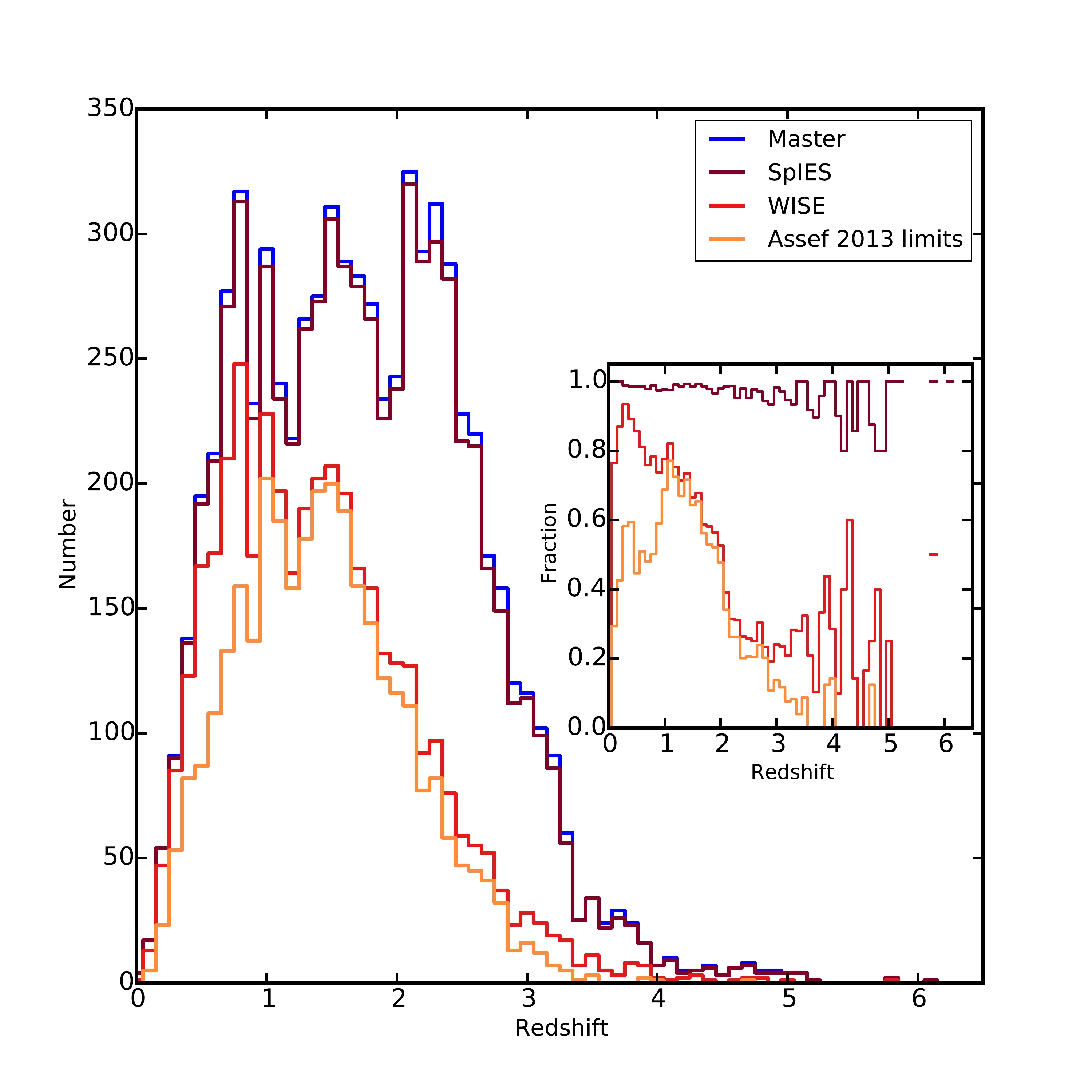}
 \includegraphics[width=9cm, height=9cm, trim=5mm 5mm 5mm 5mm, clip]{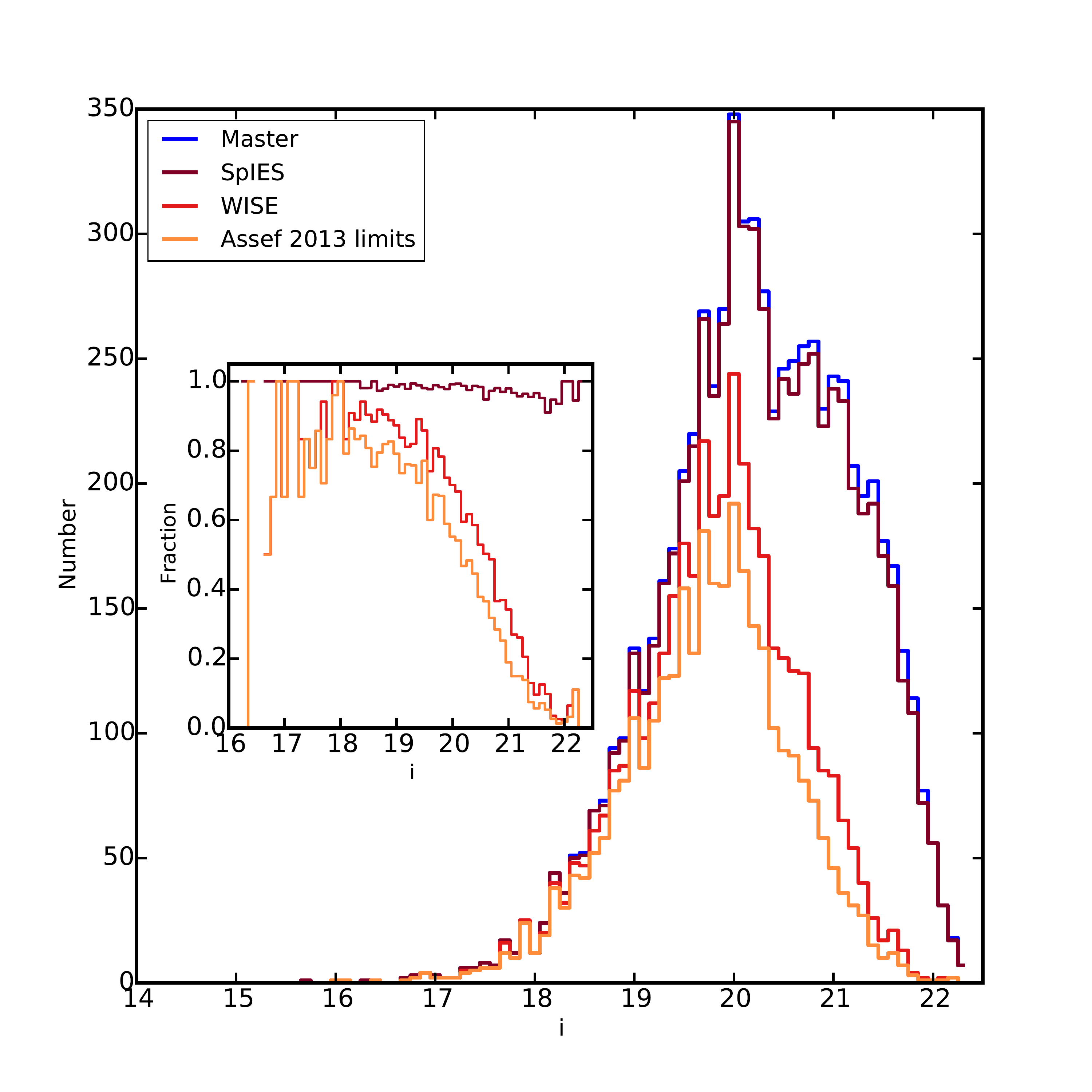}
\caption{Top: Number counts of confirmed quasar redshifts from the optical samples (blue line) in the \citet{Richards2015} ``master" catalog, the high-redshift quasars catalog of \citet{McGreer2013}, and the SDSS DR12 quasar catalog (P\^{a}ris et al. 2016, in preparation). We overplot the redshift distribution of the matched SpIES objects (dark red) and the WISE objects (red) along with the WISE data after applying the \citet{Assef2013} constraints (orange). The number counts have been enhanced by a factor of 5 at $z \ge$ 3.5 to emphasize the detections at high redshift. Bottom: The same sample of quasars, using the $i$-band magnitude as a depth comparison. The inset on both panels is the fraction of objects recovered for SpIES (dark red), WISE (red), and the \citet{Assef2013} objects (orange) with respect to the optical sample.}
\label{fig:GTR2015}
\end{center}
\end{figure}

Discovery of such objects is beyond the scope of this paper, but we show here that the SpIES data are capable of recovering such objects and have a greater ability to do so than can be achieved with the shallower {\em WISE} data.  Figure \ref{fig:GTR2015} shows redshift and $i$-band magnitude histograms of sources using the ``master" quasar catalog from \citet{Richards2015} as before. \emph{WISE} only recovers 55$\%$ of the quasars in this sample, while SpIES has superior resolution and is sufficiently deep to recover 98$\%$, including objects as faint as 22nd magnitude ($i$-band) and redshifts as high as 6. As one of the key science goals of the SpIES program is the discovery of faint, high-redshift quasars, we note that SpIES recovers 94$\%$ of these quasars with $z\ge 3.5$ as opposed to the 25$\%$ recovered by the \emph{WISE} data, and 3.5$\%$ recovered after applying the \citet{Assef2013} color cuts. 

\subsection{Summary}

The \emph{Spitzer} IRAC Equatorial Survey is supplying large-area, mid-infrared imaging of the Sloan Digital Sky Survey field Stripe 82. Utilizing mapping mode with `warm' IRAC, SpIES covers a total of $\sim$115 \sqdeg of S82 (where there is $\sim$100 \sqdeg of coverage in both bands) over two epochs, and overlaps with a wealth of ancillary data at almost every wavelength. We present the initial source catalogs for SpIES. First, a dual-band catalog containing detections in both 3.6\,$\mu$m and 4.5\,$\mu$m. Second, a 3.6\,$\mu$m-only detected catalog and, third, a 4.5\,$\mu$m-only detected catalog. In these catalogs, we report positional and photometric information, photometric errors (see Section \ref{sec:errors}), and a number of flags which are used to distinguish the high-reliability sources. The structure and analysis of these catalogs are as follows:

\begin{itemize}

\item We detect $\sim$11.6 million sources at 3.6\,$\mu$m and $\sim$12.1 million sources at 4.5\,$\mu$m, $\sim$5.4 million of which are matched between the two bands and are presented in the dual-band catalog. The remaining $\sim$6.1 million sources at 3.6\,$\mu$m and $\sim$6.6 million sources at 4.5\,$\mu$m that do not match are retained in the respective single-band only catalogs. $\sim$1.4, $\sim$3.9, and $\sim$1.4 million of these sources (3.6\,$\mu$m-only, dual-band, 4.5$\mu$m-only) are considered reliable (i.e,\ HIGH\_REL$\textgreater$0 and S/N$\textgreater$3). Much of our data analysis was performed on the dual-band catalog since it contains the most reliable sources in the survey. 

\item Using the objects in the dual-band catalog, we measured the positional accuracy (Figure \ref{fig:astrometry}) of the SpIES detections against point sources from SDSS, and have corrected the positions in the three catalogs for the measured offset. The standard deviation of this distribution is $0\farcs0008$ in RA and $0\farcs0006$ in DEC.

\item A Monte Carlo estimate of the completeness is given in Figure \ref{fig:Completeness}, which shows that SpIES is $90\%$ complete at AB magnitudes of 21.75 (7.2 $\mu$Jy) and 21.90 (6.3 $\mu$Jy) at 3.6\,$\mu$m and 4.5\,$\mu$m, respectively. Additionally, the SpIES number counts are compared with those from previous \emph{Spitzer} surveys (Figure \ref{fig:Flux_hists_spies_5s}) which, along with completeness, can be used as a measure of the survey depth. 

\item An extensive discussion of the depth is given in Section \ref{sec:depth} where we compare some of the different methods typically used to measure depth. We show that SpIES has a calculated 5$\sigma$ depth of $\sim$6.15 $\mu$Jy and $\sim$7.2 $\mu$Jy and an empirical 5$\sigma$ depth from Figure \ref{Depthfig} of  $\sim$6.13 $\mu$Jy and $\sim$5.75 $\mu$Jy at 3.6\,$\mu$m and 4.5\,$\mu$m respectively. We report the completeness and depth measurements in Table \ref{Complevel}. 

\item One of the mission goals of SpIES was to be deep enough to detect high-redshift quasars. To test how well SpIES detects these objects, we first examined the colors of different objects in the mid-infrared in Figure \ref{fig:colmag_5s}, and show that SpIES has the capability to detect these high-redshift quasars from the overlap of their mid-infrared colors. From this plot we also see that SpIES detects objects much fainter than the majority of spectroscopically confirmed high-redshift quasars. Finally, the SpIES data were matched to the known quasars in the \citet{Richards2015} ``master" quasar catalog and we show that SpIES detects a high percentage of quasars compared to \emph{WISE}, particularly at $z\ge 3.5$ (Figure \ref{fig:GTR2015}). 

\end{itemize}

The raw imaging data is available on the SHA website, and we now release the mosaics created by the SpIES team and our three detection catalogs for public use (see Appendix \ref{AppendA}).

\acknowledgements
For this research, \href{http://www.astropy.org/}{Astropy}\footnote{astropy.org} \citep{astropy}, \href{http://www.star.bristol.ac.uk/~mbt/topcat/}{TOPCAT}\footnote{starlink.ac.uk/topcat} \citep{TOPCAT}, and \href{http://www.star.bristol.ac.uk/~mbt/stilts/}{STILTS}\footnote{starlink.ac.uk/stilts} \citep{STILTS} were used for table generation and manipulation. The figures in this paper were made using \href{http://matplotlib.org/}{matplotlib}\footnote{matplotlib.org} \citep{matplotlib}, and Figures~\ref{fig:maskrad}, \ref{fig:astrometry}, \ref{Depthfig}, and \ref{fig:colmag_5s} were made with the \href{https://github.com/CKrawczyk/densityplot}{densityplot}\footnote{github.com/CKrawczyk/densityplot} \citep{densityplot} package.

This work is based [in part] on observations made with the \emph{Spitzer} Space Telescope, which is operated by the Jet Propulsion Laboratory, California Institute of Technology under a contract with NASA. Support for this work was provided by NASA through an award issued by JPL/Caltech. We would like to thank Rick Arendt, who computed the Galactic star counts for the SpIES field shown in Figure \ref{fig:Flux_hists_spies_5s}, and Matt Ashby, with whom we consulted about the SpIES number counts and depth. We acknowledge support from
CONICYT-Chile grants, 
Basal-CATA PFB-06/2007 (FEB),
FONDECYT Regular 1141218 (FEB), 
``EMBIGGEN" Anillo ACT1101 (FEB), the Ministry of Economy, Development, and Tourism's Millennium Science Initiative through grant IC120009, awarded to The Millennium Institute of Astrophysics, MAS (FEB), and NASA grant AR3-14015X and the V.M. Willaman Endowment (WNB). NPR acknowledges support from the STFC and the Ernest Rutherford Fellowship scheme.

Funding for the SDSS and SDSS-II has been provided by the Alfred P. Sloan Foundation, the Participating Institutions, the National Science Foundation, the U.S. Department of Energy, the National Aeronautics and Space Administration, the Japanese Monbukagakusho, the Max Planck Society, and the Higher Education Funding Council for England. The SDSS Web Site is \url{http://www.sdss.org/}. Funding for SDSS-III has been provided by the Alfred P. Sloan Foundation, the Participating Institutions, the National Science Foundation, and the U.S. Department of Energy Office of Science. The SDSS-III web site is \url{http://www.sdss3.org/}.
{\emph{Facilities}}: {\emph{Spitzer}} (IRAC), Sloan

\clearpage
\newpage
\appendix
\section{\\How to Access the Raw Data, Image and Catalogs} \label{AppendA}
    \subsection{Raw Data}
The raw data for SpIES can be found on the \Spitzer Heritage Archive website \url{http://sha.ipac.caltech.edu/}, where the user can input the SpIES program number (90045) and select the data type (BCD image, pBCD image, AOR).

    \subsection{Catalogs and Images}
The three detection catalogs and all of the images created by the SpIES team can be found at \url{http://www.physics.drexel.edu/~gtr/spies/}. These files have been compressed for convenience.

\newpage
\appendix
\section{\\The SpIES Astronomical Observation Requests} \label{AppendB}
\LongTables
\begin{deluxetable*}{rccrrccc}[H]
\tablecolumns{8}
\tablewidth{0pt}
\tablecaption{Full SpIES AOR list \label{AllAORs}}
\tablehead{
\colhead{Number} & \colhead{SpIES AOR Label} & \colhead{AOR Key} & \colhead{RA (J2000)}& \colhead{DEC (J2000)}& \colhead{Obs. Start}& \colhead{Obs. End}& \colhead{Integration time}\\
\colhead{} & \colhead{(IRACPC-SPIES-)} & \colhead{} & \colhead{(degrees)}& \colhead{(degrees)}& \colhead{}& \colhead{}& \colhead{(s)}}
\startdata
  1   &  22634   & 46949888  & 331.760   &   0.008333   &	2013-01-15 12:22:18   &   2013-01-15 15:36:43      & 11945 	\\
  2   &  22647   & 46945024  & 331.812   &   $-$0.008333  &	2013-01-14 20:03:50   &   2013-01-14 23:18:21      & 11956 	\\
  3   &  22915   & 46973952  & 332.275   &   0.008333	 & 	2014-09-06 22:29:12   &   2014-09-07 04:59:07	   & 23788 	\\
  4   &  22927   & 46970880  & 332.327   &   $-$0.008333	 & 	2014-09-06 11:07:02   &   2014-09-06 17:36:43	   & 23772 	\\
  5   &  221156  & 46943744  & 332.984   &   0.008333	 & 	2014-09-07 10:38:37   &   2014-09-07 15:50:27	   & 19075 	\\
  6   &  22128.  & 46955520  & 333.036   &   $-$0.008333	 & 	2014-09-07 05:25:54   &   2014-09-07 10:37:40	   & 19069 	\\
  7   &  221436  & 46969856  & 333.654   &   0.008333	 & 	2014-09-07 21:12:36   &   2014-09-08 02:24:23	   & 19072 	\\
  8   &  221449  & 46966784  & 333.706   &   $-$0.008333	 & 	2014-09-07 16:00:01   &   2014-09-07 21:11:40	   & 19064 	\\
  9   &  221717  & 46955264  & 334.324   &   0.008333	 & 	2014-09-08 16:45:17   &   2014-09-08 21:57:09	   & 19075 	\\
 10   &  221730  & 46950144  & 334.376   &   $-$0.008333	 & 	2014-09-08 03:03:36   &   2014-09-08 08:15:14	   & 19061 	\\
 11   &  221958  & 46980864  & 335.032   &   0.008333	 & 	2014-09-13 16:18:01   &   2014-09-13 22:10:02	   & 21503 	\\
 12   &  222011  & 46947584  & 335.084   &   $-$0.008333	 & 	2014-09-13 10:24:42   &   2014-09-13 16:16:39	   & 21500 	\\
 13   &  222239  & 46977792  & 335.664   &   0.008333   &	2013-01-15 17:50:48   &   2013-01-15 23:03:16      & 19071 	\\
 14   &  222251  & 46972416  & 335.716   &   $-$0.008333  &	2013-01-15 06:51:52   &   2013-01-15 12:04:24      & 19079 	\\
 15   &  222520  & 46962176  & 336.334   &   0.008333   &	2013-01-15 01:38:25   &   2013-01-15 06:51:06      & 19091 	\\
 16   &  222532  & 46958080  & 336.386   &   $-$0.008333  &	2013-01-14 09:22:06   &   2013-01-14 14:34:48      & 19101 	\\
 17   &  22280.  & 46948608  & 337.004   &   0.008333   &	2013-01-14 04:08:28   &   2013-01-14 09:21:20      & 19111 	\\
 18   &  222813  & 46943488  & 337.056   &   $-$0.008333  &	2013-01-13 22:45:08   &   2013-01-14 03:58:33      & 19115 	\\
 19   &  223041  & 46945536  & 337.674   &   0.008333   &	2013-01-13 17:30:51   &   2013-01-13 22:44:22      & 19127 	\\
 20   &  223054  & 46942208  & 337.726   &   $-$0.008333  &	2013-01-13 12:16:22   &   2013-01-13 17:29:57      & 19131 	\\
 21   &  223322  & 46971392  & 338.344   &   0.008333   &	2013-01-13 06:52:39   &   2013-01-13 12:06:22      & 19143 	\\
 22   &  223335  & 46968064  & 338.396   &   $-$0.008333  &	2013-01-13 01:37:58   &   2013-01-13 06:51:45      & 19147 	\\
 23   &  22363.  & 46948096  & 339.014   &   0.008333	 &	2014-09-14 21:53:31   &   2014-09-15 04:24:01	   & 23819 	\\
 24   &  223615  & 46953984  & 339.066   &   $-$0.008333	 &	2014-09-14 15:22:11   &   2014-09-14 21:52:32	   & 23810 	\\
 25   &  223844  & 46942464  & 339.684   &   0.008333   &	2013-01-20 02:13:16   &   2013-01-20 07:25:31      & 19065 	\\
 26   &  223856  & 46978816  & 339.736   &   $-$0.008333  &	2013-01-19 15:34:17   &   2013-01-19 20:46:34      & 19072 	\\
 27   &  224124  & 46979584  & 340.354   &   0.008333	 & 	2014-09-17 15:56:43   &   2014-09-17 21:48:42	   & 21487 	\\
 28   &  224137  & 46976000  & 340.406   &   $-$0.008333	 &	2014-09-17 10:03:25   &   2014-09-17 15:55:19	   & 21483 	\\
 29   &  22445.  & 46951680  & 341.024   &   0.008333   &	2013-01-18 22:36:06   &   2013-01-19 03:49:14      & 19102 	\\
 30   &  224418  & 46962432  & 341.076   &   $-$0.008333  &	2013-01-18 17:22:04   &   2013-01-18 22:35:12      & 19106 	\\
 31   &  224646  & 46952448  & 341.694   &   0.008333   &	2013-01-20 12:50:36   &   2013-01-20 18:03:20      & 19089 	\\
 32   &  224659  & 46949376  & 341.746   &   $-$0.008333  &	2013-01-20 07:26:56   &   2013-01-20 12:39:42      & 19093 	\\
 33   &  224927  & 46976512  & 342.364   &   0.008333   &	2013-01-22 12:04:10   &   2013-01-22 17:16:26      & 19068 	\\
 34   &  224939  & 46973184  & 342.416   &   $-$0.008333  &	2013-01-22 06:42:00   &   2013-01-22 11:54:18      & 19072 	\\
 35   &  22528.  & 46951168  & 343.034   &   0.008333   &	2013-01-18 06:42:57   &   2013-01-18 11:56:32      & 19139 	\\
 36   &  225220  & 46969600  & 343.086   &   $-$0.008333  &	2013-01-18 12:07:13   &   2013-01-18 17:20:49      & 19136 	\\
 37   &  225448  & 46960896  & 343.704   &   0.008333   &	2013-01-19 04:56:31   &   2013-01-19 10:09:46      & 19135 	\\
 38   &  22551.  & 46952192  & 343.756   &   $-$0.008333  &	2013-01-19 10:10:41   &   2013-01-19 15:23:53      & 19132 	\\
 39   &  225729  & 46946560  & 344.374   &   0.008333   &	2013-01-22 01:28:02   &   2013-01-22 06:40:47      & 19104 	\\
 40   &  225742  & 46980352  & 344.426   &   $-$0.008333  &	2013-01-21 13:15:52   &   2013-01-21 18:28:43      & 19111 	\\
 41   &  23010   & 46967040  & 345.044   &   0.008333	 & 	2014-09-22 01:15:20   &   2014-09-22 07:07:07	   & 21486 	\\
 42   &  23023   & 46963712  & 345.096   &   $-$0.008333	 & 	2014-09-22 07:08:28   &   2014-09-22 13:00:19	   & 21489 	\\
 43   &  23251   & 46953472  & 345.714   &   0.008333   &	2013-01-20 18:05:04   &   2013-01-20 23:18:42      & 19140 	\\
 44   &  2333.   & 46978304  & 345.766   &   $-$0.008333  &	2013-01-20 23:35:28   &   2013-01-21 04:49:06      & 19138 	\\
 45   &  23532   & 46977280  & 346.345   &   0.008333 	 & 	2014-09-19 05:22:35   &   2014-09-19 11:13:11	   & 21404 	\\
 46   &  23544   & 46974464  & 346.397   &   $-$0.008333	 & 	2014-09-19 11:14:33   &   2014-09-19 17:05:15	   & 21409 	\\
 47   &  23812   & 46963968  & 347.054   &   0.008333 	 & 	2014-09-20 05:39:13   &   2014-09-20 10:50:41	   & 19061 	\\
 48   &  23825   & 46959360  & 347.106   &   $-$0.008333	 & 	2014-09-20 10:51:35   &   2014-09-20 16:03:08	   & 19064 	\\
 49   &  231053  & 46954752  & 347.724   &   0.008333 	 & 	2014-09-21 10:03:17   &   2014-09-21 15:15:05	   & 19075 	\\
 50   &  23116.  & 46971136  & 347.776   &   $-$0.008333	 & 	2014-09-21 15:15:59   &   2014-09-21 20:27:49	   & 19076 	\\
 51   &  231334  & 46978560  & 348.394   &   0.008333 	 & 	2014-09-24 03:00:32   &   2014-09-24 08:13:02	   & 19104 	\\
 52   &  231347  & 46975232  & 348.446   &   $-$0.008333	 & 	2014-09-23 21:47:10   &   2014-09-24 02:59:36	   & 19101 	\\
 53   &  231615  & 46964480  & 349.064   &   0.008333 	 & 	2014-09-24 08:23:13   &   2014-09-24 13:35:39	   & 19097 	\\
 54   &  231627  & 46960640  & 349.116   &   $-$0.008333	 & 	2014-09-24 13:36:33   &   2014-09-24 18:49:04	   & 19101 	\\
 55   &  231856  & 46951936  & 349.734   &   0.008333 	 & 	2014-09-27 06:29:48   &   2014-09-27 11:42:33	   & 19133 	\\
 56   &  23198.  & 46961664  & 349.786   &   $-$0.008333	 & 	2014-09-26 15:23:00   &   2014-09-26 20:35:32	   & 19124 	\\
 57   &  232136  & 46949120  & 350.404   &   0.008333   &	2013-01-29 18:01:35   &   2013-01-29 23:13:56      & 19078 	\\
 58   &  232149  & 46944000  & 350.456   &   $-$0.008333  &	2013-01-29 12:48:22   &   2013-01-29 18:00:42      & 19083 	\\
 59   &  232417  & 46972928  & 351.074   &   0.008333   &	2013-01-27 06:05:06   &   2013-01-27 11:18:22      & 19121 	\\
 60   &  232430  & 46968320  & 351.126   &   $-$0.008333  &	2013-01-27 00:51:00   &   2013-01-27 06:04:12      & 19125 	\\
 61   &  232658  & 46959104  & 351.744   &   0.008333   &	2013-01-25 07:06:46   &   2013-01-25 12:20:32      & 19159 	\\
 62   &  232711  & 46954240  & 351.796   &   $-$0.008333  &	2013-01-24 10:56:43   &   2013-01-24 16:10:41      & 19172 	\\
 63   &  232939  & 46944512  & 352.414   &   0.008333   &	2013-01-27 16:42:59   &   2013-01-27 21:56:29      & 19134 	\\
 64   &  232951  & 46979072  & 352.466   &   $-$0.008333  &	2013-01-27 11:28:32   &   2013-01-27 16:42:05      & 19138 	\\
 65   &  233220  & 46979840  & 353.084   &   0.008333   &	2013-01-25 17:44:54   &   2013-01-25 22:58:59      & 19171 	\\
 66   &  233232  & 46976256  & 353.136   &   $-$0.008333  &	2013-01-25 12:21:49   &   2013-01-25 17:35:53      & 19176 	\\
 67   &  23350.  & 46966272  & 353.754   &   0.008333   &	2013-09-24 20:07:15   &   2013-09-25 01:20:29      & 19129 	\\
 68   &  233513  & 46962944  & 353.806   &   $-$0.008333  &	2013-09-24 14:53:12   &   2013-09-24 20:06:19      & 19126 	\\
 69   &  233741  & 46952960  & 354.424   &   0.008333   &	2013-09-27 05:28:49   &   2013-09-27 10:42:24      & 19155 	\\
 70   &  233754  & 46949632  & 354.476   &   $-$0.008333  &	2013-09-26 23:28:02   &   2013-09-27 04:41:30      & 19150 	\\
 71   &  234022  & 46950656  & 355.094   &   0.008333   &	2013-09-26 18:13:41   &   2013-09-26 23:26:56      & 19138 	\\
 72   &  234035  & 46945792  & 355.146   &   $-$0.008333  &	2013-09-26 07:33:02   &   2013-09-26 12:46:05      & 19131 	\\
 73   &  23433.  & 46965760  & 355.764   &   0.008333   &	2013-09-23 23:03:13   &   2013-09-24 04:16:06      & 19092 	\\
 74   &  234315  & 46971648  & 355.816   &   $-$0.008333  &	2013-09-23 02:21:37   &   2013-09-23 07:34:09      & 19076 	\\
 75   &  234544  & 46961152  & 356.434   &   0.008333   &	2013-09-24 09:39:16   &   2013-09-24 14:51:42      & 19089 	\\
 76   &  234556  & 46957568  & 356.486   &   $-$0.008333  &	2013-09-24 04:16:58   &   2013-09-24 09:29:46      & 19086 	\\
 77   &  234824  & 46947072  & 357.104   &   0.008333   &	2013-09-27 15:57:49   &   2013-09-27 21:11:02      & 19124 	\\
 78   &  234837  & 46943232  & 357.156   &   $-$0.008333  &	2013-09-27 10:43:48   &   2013-09-27 15:56:53      & 19120 	\\
 79   &  23515.  & 46969088  & 357.774   &   0.008333 	 & 	2014-10-01 09:05:27   &   2014-10-01 14:17:10	   & 19081 	\\
 80   &  235118  & 46980096  & 357.826   &   $-$0.008333	 & 	2014-10-01 14:18:04   &   2014-10-01 19:29:53	   & 19083 	\\
 81   &  235346  & 46970624  & 358.444   &   0.008333 	 & 	2014-10-01 03:44:12   &   2014-10-01 08:55:37	   & 19062 	\\
 82   &  235359  & 46967808  & 358.496   &   $-$0.008333	 & 	2014-09-30 22:32:02   &   2014-10-01 03:43:17	   & 19052 	\\
 83   &  235627  & 46956288  & 359.114   &   0.008333 	 &	2014-10-06 05:30:49   &   2014-10-06 10:43:44	   & 19132 	\\
 84   &  235639  & 46953216  & 359.166   &   $-$0.008333	 & 	2014-10-06 00:17:04   &   2014-10-06 05:29:53	   & 19130 	\\
 85   &  23598.  & 46956800  & 359.784   &   0.008333 	 & 	2014-10-05 18:54:42   &   2014-10-06 00:07:17	   & 19119 	\\
 86   &  235920  & 46974976  & 359.836   &   $-$0.008333	 & 	2014-10-05 08:23:09   &   2014-10-05 13:35:37	   & 19112 	\\
 87   &  0148    & 46974208  & 0.454     &   0.008333   &	2013-02-01 07:42:52   &   2013-02-01 12:56:56      & 19178 	\\
 88   &  021.    & 46964736  & 0.506     &   $-$0.008333  &	2013-02-01 02:27:55   &   2013-02-01 07:41:58      & 19184 	\\
 89   &  0429    & 46958848  & 1.124     &   0.008333   &	2013-09-30 00:07:47   &   2013-09-30 05:20:39      & 19103 	\\
 90   &  0442    & 46954496  & 1.176     &   $-$0.008333  &	2013-09-30 05:21:33   &   2013-09-30 10:34:33      & 19106 	\\
 91   &  0710    & 46942720  & 1.794     &   0.008333   &	2013-09-30 16:20:56   &   2013-09-30 21:33:54      & 19103 	\\
 92   &  0723    & 46978048  & 1.846     &   $-$0.008333  &	2013-09-30 21:43:31   &   2013-10-01 02:56:33      & 19106 	\\
 93   &  0951    & 46969344  & 2.464     &   0.008333   &	2013-10-01 02:57:20   &   2013-10-01 08:10:19      & 19101 	\\
 94   &  0103.   & 46965504  & 2.516     &   $-$0.008333  &	2013-10-01 19:30:29   &   2013-10-02 00:43:36      & 19108 	\\
 95   &  01232   & 46964224  & 3.134     &   0.008333   &	2013-10-02 00:53:18   &   2013-10-02 06:06:01      & 19104 	\\
 96   &  01244   & 46960384  & 3.186     &   $-$0.008333  &	2013-10-02 07:03:50   &   2013-10-02 12:16:38      & 19107 	\\
 97   &  01512   & 46950912  & 3.804     &   0.008333   &	2013-10-03 01:26:08   &   2013-10-03 06:39:09      & 19108 	\\
 98   &  01525   & 46946304  & 3.856     &   $-$0.008333  &	2013-10-03 19:40:08   &   2013-10-04 00:53:17      & 19119 	\\
 99   &  01753   & 46975488  & 4.474     &   0.008333   &	2013-10-04 22:50:39   &   2013-10-05 04:03:46      & 19127 	\\
100   &  0186.   & 46953728  & 4.526     &   $-$0.008333  &	2013-10-05 10:58:43   &   2013-10-05 16:11:57      & 19132 	\\
101   &  02034   & 46972672  & 5.144     &   0.008333   &	2013-10-05 16:12:44   &   2013-10-05 21:25:58      & 19127 	\\
102   &  02047   & 46970112  & 5.196     &   $-$0.008333  &	2013-10-05 21:35:40   &   2013-10-06 02:49:00      & 19129 	\\
103   &  02315   & 46958336  & 5.814     &   0.008333 	 & 	2014-10-08 18:12:29   &   2014-10-08 23:24:09	   & 19074 	\\
104   &  02327   & 46955776  & 5.866     &   $-$0.008333	 & 	2014-10-08 12:50:34   &   2014-10-08 18:02:28	   & 19064 	\\
105   &  02556   & 46944256  & 6.484     &   0.008333 	 & 	2014-10-09 06:57:36   &   2014-10-09 12:09:18	   & 19072 	\\
106   &  0268.   & 46956544  & 6.536     &   $-$0.008333	 & 	2014-10-09 01:45:09   &   2014-10-09 06:56:41	   & 19064 	\\
107   &  02836   & 46970368  & 7.154     &   0.008333 	 & 	2014-10-12 21:56:43   &   2014-10-13 03:09:07	   & 19118 	\\
108   &  02849   & 46967552  & 7.206     &   $-$0.008333	 & 	2014-10-13 03:19:55   &   2014-10-13 08:32:24	   & 19120 	\\
109   &  03117   & 46968576  & 7.824     &   0.008333 	 & 	2014-10-13 08:33:11   &   2014-10-13 13:45:38	   & 19115 	\\
110   &  03130   & 46962688  & 7.876     &   $-$0.008333	 & 	2014-10-13 13:46:32   &   2014-10-13 18:59:06	   & 19118 	\\
111   &  03358   & 46955008  & 8.494     &   0.008333 	 & 	2014-10-15 02:56:10   &   2014-10-15 08:09:03	   & 19130 	\\
112   &  03411   & 46948352  & 8.546     &   $-$0.008333	 & 	2014-10-14 21:33:39   &   2014-10-15 02:46:29	   & 19126 	\\
113   &  03639   & 46979328  & 9.164     &   0.008333 	 & 	2014-10-13 21:13:54   &   2014-10-14 02:26:13	   & 19102 	\\
114   &  03651   & 46973440  & 9.216     &   $-$0.008333	 & 	2014-10-14 02:27:07   &   2014-10-14 07:39:31	   & 19103 	\\
115   &  03920   & 46963200  & 9.834     &   0.008333 	 & 	2014-10-15 08:10:08   &   2014-10-15 13:22:49	   & 19115 	\\
116   &  03932   & 46958592  & 9.886     &   $-$0.008333	 & 	2014-10-15 19:29:13   &   2014-10-16 00:42:02	   & 19120 	\\
117   &  0420.   & 46960128  & 10.504    &   0.008333 	 & 	2014-10-16 06:05:16   &   2014-10-16 11:17:42	   & 19119 	\\
118   &  04213   & 46957056  & 10.556    &   $-$0.008333	 & 	2014-10-16 00:42:52   &   2014-10-16 05:55:36	   & 19115 	\\
119   &  04441   & 46946816  & 11.174    &   0.008333 	 & 	2014-10-16 17:26:02   &   2014-10-16 22:38:30	   & 19117 	\\
120   &  04454   & 46942976  & 11.226    &   $-$0.008333	 & 	2014-10-16 22:39:24   &   2014-10-17 03:51:59	   & 19119 	\\
121   &  04722   & 46971904  & 11.844    &   0.008333	 &	2014-10-18 16:32:54   &	  2014-10-18 21:45:50	   & 19133 	\\
122   &  04735   & 46968832  & 11.896    &   $-$0.008333  &	2014-10-18 05:01:26   &   2014-10-18 10:14:13	   & 19126 	\\
123   &  0503.   & 46959872  & 12.514    &   0.008333	 &	2014-10-19 03:09:25   &   2014-10-19 08:22:23	   & 19130 	\\
124   &  05015   & 46966016  & 12.566    &   $-$0.008333  &	2014-10-18 21:46:41   &   2014-10-19 02:59:33	   & 19127 	\\
125   &  05244   & 46956032  & 13.184    &   0.008333	 &	2014-10-17 23:34:29   &   2014-10-18 04:46:50 	   & 19103 	\\
126   &  05256   & 46952704  & 13.236    &   $-$0.008333  &	2014-10-17 18:21:21   &   2014-10-17 23:33:34	   & 19098 	\\
127   &  05524   & 46980608  & 13.854    &   0.008333	 &	2014-10-20 21:34:09   &   2014-10-21 02:47:05	   & 19136 	\\
128   &  05537   & 46976768  & 13.906    &   $-$0.008333  &	2014-10-20 16:20:23   &   2014-10-20 21:33:14	   & 19133 	\\
129   &  1490    & 46967296  & 27.254    &   0.008333   &	2013-10-18 17:40:42   &   2013-10-18 22:51:25      & 18965 	\\
130   &  14913   & 46963456  & 27.306    &   $-$0.008333  &	2013-10-18 22:52:20   &   2013-10-19 04:03:05      & 18970 	\\
131   &  15141   & 46965248  & 27.924    &   0.008333   &	2013-10-19 11:07:27   &   2013-10-19 16:18:12      & 18968 	\\
132   &  15154   & 46961408  & 27.976    &   $-$0.008333  &	2013-10-19 16:19:07   &   2013-10-19 21:29:53      & 18970 	\\
133   &  15422   & 46951424  & 28.594    &   0.008333   &	2013-10-20 13:16:13   &   2013-10-20 18:27:12      & 18975 	\\
134   &  15435   & 46947328  & 28.646    &   $-$0.008333  &	2013-10-20 18:37:01   &   2013-10-20 23:47:36      & 18977 	\\
135   &  1573.   & 46966528  & 29.264    &   0.008333   &	2013-10-20 23:48:22   &   2013-10-21 04:58:50      & 18970 	\\
136   &  15715   & 46972160  & 29.316    &   $-$0.008333  &	2013-10-21 10:44:00   &   2013-10-21 15:54:36      & 18977 	\\
137   &  15944   & 46961920  & 29.934    &   0.008333   &	2013-10-21 15:55:22   &   2013-10-21 21:05:52      & 18970 	\\
138   &  15956   & 46957824  & 29.986    &   $-$0.008333  &	2013-10-21 21:06:45   &   2013-10-22 02:17:18      & 18971 	\\
139   &  2224    & 46977536  & 30.604    &   0.008333   &	2013-10-24 13:46:24   &   2013-10-24 18:57:49      & 19021 	\\
140   &  2237    & 46974720  & 30.656    &   $-$0.008333  &	2013-10-24 18:58:43   &   2013-10-25 00:10:09      & 19022 	\\
141   &  255.    & 46948864  & 31.274    &   0.008333   &	2013-10-25 00:20:14   &   2013-10-25 05:31:32      & 19014 	\\
142   &  2518    & 46959616  & 31.326    &   $-$0.008333  &	2013-10-25 08:39:48   &   2013-10-25 13:51:12      & 19021 	\\
143   &  2746    & 46950400  & 31.944    &   0.008333   &	2013-10-27 01:14:33   &   2013-10-27 06:26:30      & 19045 	\\
144   &  2759    & 46946048  & 31.996    &   $-$0.008333  &	2013-10-27 11:23:55   &   2013-10-27 16:35:58      & 19052 	\\
145   &  21027   & 46944768  & 32.614    &   0.008333   &	2013-10-28 18:25:44   &   2013-10-28 23:37:49      & 19059 	\\
146   &  21039   & 46981120  & 32.666    &   $-$0.008333  &	2013-10-28 23:38:43   &   2013-10-29 04:50:49      & 19061 	\\
147   &  2138.   & 46945280  & 33.284    &   0.008333   &	2013-10-31 13:24:25   &   2013-10-31 18:37:17      & 19090 	\\
148   &  21320   & 46964992  & 33.336    &   $-$0.008333  &	2013-10-31 18:47:26   &   2013-11-01 00:00:17      & 19093 	\\
149   &  21548   & 46957312  & 33.954    &   0.008333   &	2013-11-01 18:29:22   &   2013-11-01 23:42:19      & 19098 	\\
150   &  2161.   & 46947840  & 34.006    &   $-$0.008333  &	2013-11-01 23:43:13   &   2013-11-02 04:56:10      & 19100 	\\
151   &  21829   & 46941952  & 34.624    &   0.008333	 &	2014-11-08 22:10:40   &   2014-11-09 03:23:03	   & 18801 	\\
152   &  21842   & 46975744  & 34.676    &   $-$0.008333	 &	2014-11-08 16:48:59   &   2014-11-08 22:01:17	   & 18801 	\\
153   &  22110   & 46977024  & 35.294    &   0.008333	 &	2014-11-08 11:35:49   &   2014-11-08 16:47:54	   & 18802 	\\
154   &  22123   & 46973696  & 35.346    &   $-$0.008333	 &	2014-11-08 06:14:41   &   2014-11-08 11:26:35	   & 18801 
\enddata
\tablecomments{This table shows the single epoch AORs in the SpIES survey. Each set of 2 AORs overlap approximately the same area, for instance, AOR number 1 and 2 overlap with each other to image one region on the sky.}
\end{deluxetable*}

\bibliographystyle{yahapj}

\bibliography{ms}

\end{document}